\documentclass[aps,prb,
preprint,
showpacs,
floatfix,
]{revtex4-1}
\usepackage{bm}
\usepackage{amsmath}    
\usepackage{graphicx}   
\usepackage[hang]{subfigure}
\newcommand{\jun}{junction }
\newcommand{\juns}{junctions }
\newcommand{\Jos}{Josephson }
\newcommand{\elli}{elliptic }
\newcommand{\ann}{annular }

\begin{document}
\title{Elliptic Annular \Jos Tunnel Junctions in an external magnetic field: The statics}
\author{Roberto Monaco}
\email{r.monaco@cib.na.cnr.it}
\affiliation{Consiglio Nazionale delle Ricerche, Istituto di Cibernetica, Comprensorio Olivetti, 80078 Pozzuoli, Italy}
\author{Carmine Granata}
\email{c.granata@cib.na.cnr.it}
\affiliation{Consiglio Nazionale delle Ricerche, Istituto di Cibernetica, Comprensorio Olivetti, 80078 Pozzuoli, Italy}
\author{Antonio Vettoliere}
\email{a.vettoliere@cib.na.cnr.it}
\affiliation{Consiglio Nazionale delle Ricerche, Istituto di Cibernetica, Comprensorio Olivetti, 80078 Pozzuoli, Italy}
\author{Jesper Mygind}
\email{myg@fysik.dtu.dk}
\affiliation{DTU Physics, B309, Technical University of Denmark, DK-2800 Lyngby, Denmark}

\date{\today}
 
\begin{abstract}

We have investigated the static properties of one-dimensional planar Josephson tunnel junctions in the most general case of elliptic annuli. We have analyzed the dependence of the critical current in the presence of an external magnetic field applied either in the junction plane or in the perpendicular direction. We report a detailed study of both short and long elliptic annular junctions having different eccentricities. For junctions having a normalized perimeter less than one the threshold curves are derived and computed even in the case with one trapped Josephson vortex. For longer junctions a numerical analysis is carried out after the derivation of the appropriate Perturbed sine-Gordon Equation. For a given applied field we find that a number of different phase profiles exist which differ according to the number of fluxon-antifluxon pairs. We demonstrate that in samples made by specularly symmetric electrodes a transverse magnetic field is equivalent to an in-plane field applied in the direction of the current flow. Varying the ellipse eccentricity we reproduce all known results for linear and ring-shaped Josephson tunnel junctions. Experimental data on high-quality $Nb/Al$-$AlOx/Nb$ elliptic annular junctions support the theoretical analysis provided self-field effects are taken into account.
\end{abstract}
\maketitle
\section{Introduction}
Circular \ann \Jos tunnel \juns (JTJs), consisting of two superconducting rings coupled by a thin dielectric tunneling layer were recognized to be the ideal benchmark to test both the statics and the dynamics of sine-Gordon solitons in the presence of a periodic potential\cite{gronbech,ustinov,PRB98,wallraf}. In this context, a soliton is a current vortex, also called a \Jos vortex or a fluxon, circulating around the \jun and carrying one magnetic flux quantum. A spatially periodic potential for the fluxon can be easily implemented by a uniform magnetic field applied in the plane of the annulus. However, for JTJs having a not simply-connected topology the most general and at the same time regular geometry is provided by an \elli annulus. At variance with a circle that has infinitely many axes of symmetry, an ellipse has two axes of symmetry. This geometrical symmetry breaking which comes with an associated non-uniformity of the radius of curvature has been exploited in several physical systems, e.g., to increase the focal depth\cite{welford} or the resolving power\cite{born} of annular apertures and as antenna reflectors in the microwave region\cite{kathuria}. Elliptic \ann \Jos tunnel junctions (EAJTJs) serve as an handy tool for the realization of complex periodic potentials, including those lacking spatial reflection symmetry, known as ratchet potentials\cite{magnasco}. An additional motivation to study EAJTJs is the intention to cast in one unique class many apparently different JTJ configurations, including the linear geometry commonly studied in the context of JTJs.  
 
\noindent In this article we focus our attention on the static configurations
of the phase in EAJTJs; the dynamics of solitons will be the subject of another article. Most of the work will be focused on EAJTJs having the so called \textit{Lyngby-type} geometry\cite{davidson85}, that refers to a specularly symmetric configuration in which the height of the current carrying electrodes matches one of the ellipse outer axis (e.g. see Figures~\ref{layout}(a) and (b)).  We have measured the dependence of the zero-voltage \Jos current for a large number of samples. The experimental data are compared with the analytical results and with the numerical results obtained by solving the appropriate partial differential equation.

\noindent The paper is organized as follows. In next Section we first consider a quater-elliptic \ann \jun immersed in a uniform in-plane magnetic field and derive the threshold curves for \juns having different ellipticity; later we extend the analysis to full ellipses with possible \Jos vortices trapped in the \ann barrier. In Section III we derive the appropriate partial differential equation for an electrically long EAJTJ; later we present the results of the numerical simulations and outline the effects of the magnetic field induced by the \jun bias current. Thereafter, we investigate the consequences of a magnetic field applied perpendicular to the \jun plane for different geometrical configurations; for Lyngby-type samples we establish the equivalence between a transverse magnetic field and an in-plane magnetic field applied along the flow of the bias current. Besides, we suggest a simple geometrical configuration that implements the ideal step-like deterministic periodic ratchet potential. In Section V we describe the fabrication of our $Nb/Al$-$AlOx/Nb$ samples, the different geometries that have been realized and the experimental setup; finally, we present and discuss the experimental data of long EAJTJs with both in-plane and transverse magnetic fields. The conclusions are drawn in Section VI.

\section{Short elliptic \ann junctions}

In this section we derive the equations which describe the behavior of a small EAJTJ in the presence of a uniform (static) magnetic field, $H_{\parallel}$, applied along one of its axis. In order to confer the largest generality to the analysis, we will begin by considering a quarter-elliptic planar junction which, depending on its eccentricity, will include also the cases of linear and quarter-annular\cite{Shaju2002} junctions. Later on, we will treat an EAJTJ as the parallel combination of four quarter-\elli ones subject to periodic boundary conditions.

\subsection{Quarter-elliptic junctions}

\begin{figure}[tb]
\centering
\includegraphics[width=7cm]{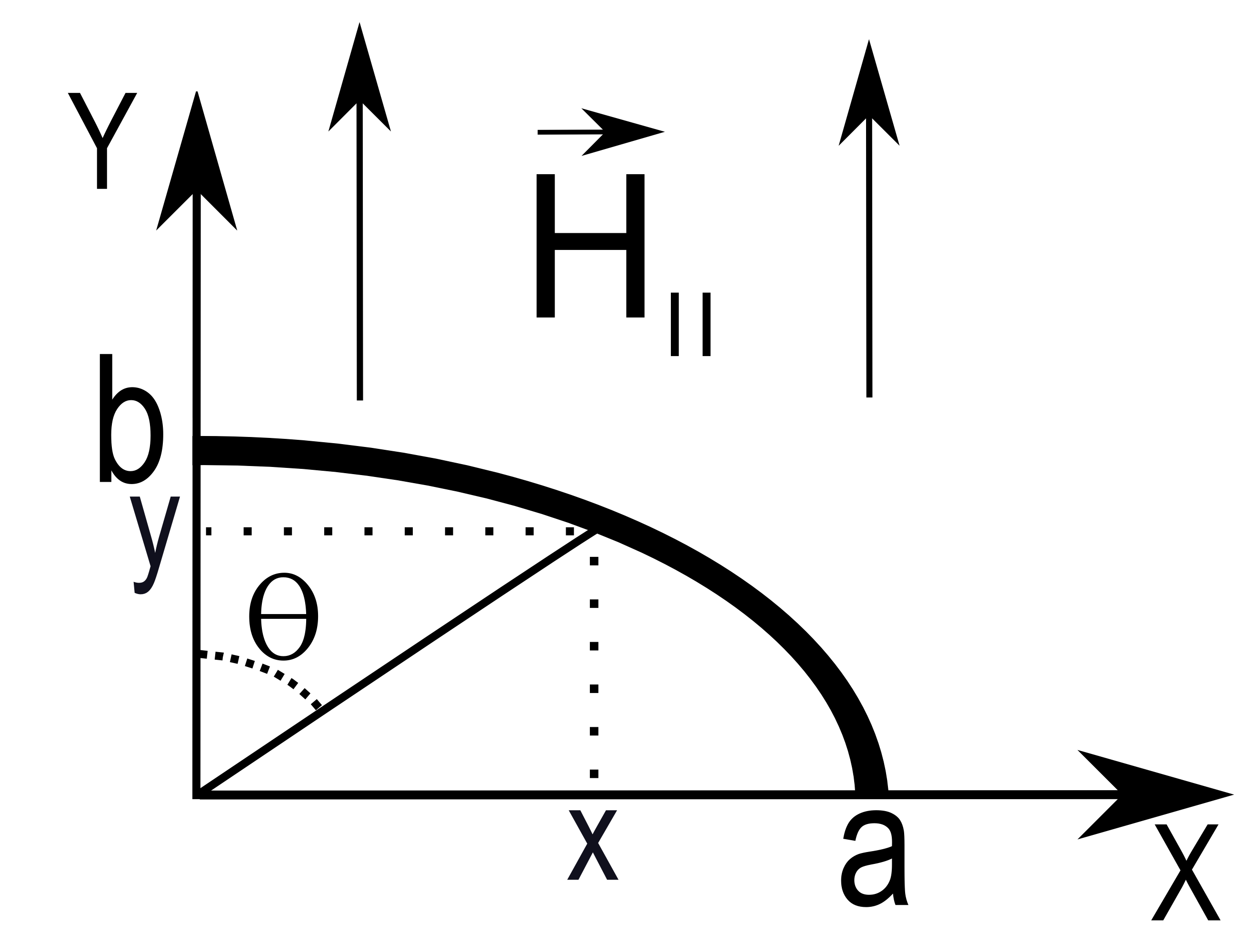}
\caption{Quarter-\elli \Jos planar tunnel \jun in a uniform in-plane magnetic field ${\bf H_{\parallel}}\equiv \left(0, H_{\parallel} \right)$.}
\label{quarterellipse}
\end{figure}

As shown in Figure~\ref{quarterellipse}, let the quarter-ellipse lay in the plan identified by the $X$-$Y$ Cartesian coordinate system whose origin coincides with the center of symmetry and whose axes are directed along the principal semi-axes $a$ and $b$ of its master ellipse. The curve is described by the the parametric equations $x=a\sin \tau$ and $y=b \cos \tau$, where $\tau$ ($0\leq \tau \leq \pi/2$) is a parameter measured clockwise from the positive $Y$-axis such that $\tau \equiv \text{ArcTan}\,\rho x/y$, not to be confused with the polar angle $\theta \equiv \text{ArcTan}\, x/y$. We defined $\rho\equiv b/a$ the axes ratio and we will also make use of the master-ellipse eccentricity $e^2 \equiv 1-\rho^2$. For a circular arc, $\rho=1$ (no eccentricity), so $\tau$ and $\theta$ coincide, while for $\rho\neq1$, $\tau=\theta$ only for $\theta= m \pi/2$. In the case of a thin circular ring with mean radius $r$, we would introduce the curvilinear coordinate $s(\theta)=r \theta$ such that $s$ linearly increases by one perimeter (circumference) as $\theta$ changes by $2 \pi$. Along an elliptic arc we instead introduce the non-linear curvilinear coordinate $s(\tau)=a \int_{0}^{\tau} \mathcal{I}(\tau')d\tau' =a \text{E}(\tau,e^2)$, where $\mathcal{I}(\tau)\equiv \sqrt{1-e^2\sin^2\!\tau} =\sqrt{\cos^2\!\tau+\rho^2\sin^2\!\tau}$ is the integrand of the \textit{incomplete} elliptic integral of the second kind, $\text{E}(\tau,e^2)$.

\noindent In Josephson's original description the quantum mechanical phase difference, $\phi$, across the barrier is related to the magnetic field, ${\bf H}$, inside the barrier\cite{brian}:
\vskip -10pt
\begin{equation}
\label{gra}{\bf \nabla} \phi =
\kappa{\bf H}\times {\bf u}_z ,
\end{equation}

\noindent in which ${\bf u}_z$ is a unit vector orthogonal to the \jun plane and $\kappa^{-1}\equiv \Phi_0/2\pi\mu_0 d_m$, where $\Phi_0$ is the magnetic flux quantum, $\mu_0$ the vacuum permeability, and $d_m= \lambda_{b} \tanh d_b /2 \lambda_{b} + \lambda_{t} \tanh d_t /2 \lambda_{t}$ is the \jun \textit{magnetic} penetration depth\cite{wei}, where $\lambda_{b,t}$ and $d_{b,t}$ are, respectively, the bulk magnetic penetration depths and thicknesses of the junction bottom and top films; $d_m$ reduces to $ \lambda_{b}+\lambda_{t}$ in the case of thick superconducting films ($d_{b,t}$ larger than $\simeq 4\lambda_{b,t}$). The subscripts $b$ and $t$ will be adopted to label the quantities which refer, respectively, to the bottom and top/wiring layers. 

\noindent ${\bf H}$ in Eq.(\ref{gra}) is the total magnetic field that, in general, is given by the sum of an externally applied field and the self-field generated by the current flowing in the junction. Eq.(\ref{gra}) states that, among other things, $\phi$ is not sensitive to fields along the $Z$-direction; this is only true in the ideal case of a bare Josephson sandwich with no electrical connections. In real devices the presence of the current carrying electrodes not only alters the effect of a barrier-parallel field, but also makes the junction sensitive to a transverse field\cite{JAP07,JAP08}. The way such field induces screening currents, which, in turn, generate in-plane magnetic fields is well understood in rectangular JTJs; in Section IV, we will discuss how a similar approach works with not simply connected junctions.

\noindent For the time being, we assume that the quarter-\elli \jun is electrically short, i.e., the arc mean length, $L$, is small compared to the Josephson penetration length $\lambda_J \equiv \sqrt{\Phi_0/ 2\pi \mu _{0}d_j J_{c}}$, where $d_j$ is the junction \textit{current} thickness\cite{wei} $d_j= \lambda_{b} \coth d_b / \lambda_{b} + \lambda_{t} \coth d_t /\lambda_{t} \geq \lambda_b+\lambda_t \geq d_m$; for thick film junctions, $d_j \simeq d_m \simeq \lambda_{b}+\lambda_{t}$. Further, we assume that the arc has a finite width, $W$, much smaller than the quarter-ellipse semi-axes. Under these conditions a spatially homogeneous field applied in the $X$-$Y$ plane fully penetrates into the barrier. In our specific case, the external field is applied perpendicular to the $a$-axis, i.e., along the positive $Y$-direction, ${\bf H_{\parallel}}\equiv \left(0, H_{\parallel} \right)$. According to Eq.(\ref{gra}), the \Jos phase only changes in the $X$-direction, i.e., $\partial \phi / \partial y=0$ and $d \phi / d x = -\kappa H_{\parallel}$; then $\phi(x)=- \kappa H_{\parallel} x + \phi_0$, where $\phi_0$ is an integration constant. The dependence on $\tau$ is:

\vskip -15pt
\begin{equation}
\label{phiditau}
\phi(\tau) = -h_{\parallel} \sin \, \tau + \phi_0,
\end{equation} 


\noindent where $h_{\parallel}$ is the strength of the in plane field, $H_{\parallel}$, normalized to $(\kappa a)^{-1} =\Phi_0/2\pi \mu_0 d_m a$. The local density, $J_J$, of the \Jos current at a point $\vec{r}$ inside the barrier area is\cite{brian} $J_J(\vec{r}) =J_c(\vec{r}) \sin \phi(\vec{r})$, where $J_c$ is the maximum \Jos current density. The \Jos current, $I_J$, through the barrier is obtained integrating $J_J$ over the junction area, $I_J=\int_A J_J dA$; in force of the one-dimensional approximation, $A=WL$ and the elementary surface element is $dA=dWdL$, where $dL=ds=a\, \mathcal{I}(\tau)d\tau$ is the elliptic elementary arc. Assuming $J_c$ uniform over the barrier area and recalling that $\phi$ is constant along the annulus width:
\vskip -10pt
\begin{equation}
\label{IJ}
I_J= J_c W \int_L \!\! \sin \phi(\tau)dL(\tau)=J_c W a \int_{0}^{\pi/2} \!\!\!\! \mathcal{I}(\tau) \sin(-h_{\parallel} \sin \, \tau + \phi_0) d\tau.
\end{equation}

\noindent If the in-plane field is applied along the $X$-direction, the factor $-h_{\parallel}\sin \, \tau$ should be replaced by $h_{\parallel} \rho \cos\, \tau$ (or, equivalently, one should operate the transformation $\rho \longrightarrow 1/\rho$). From Eq.(\ref{IJ}) with $e^2=1$ ($\rho=0$) we easily recover the Fraunhofer-like magnetic diffraction pattern (MDP), $|\sin \,h_{\parallel}/ h_{\parallel}| =|\text{Sinc}\, h_{\parallel}|$, typical of small linear junctions.

\noindent The critical current, $I_c$, of a quarter-\elli \jun can be found by maximizing\cite{barone} Eq.(\ref{IJ}) with respect to $\phi_0$:
\vskip -10pt
\begin{equation} \label{Icquarter}
I_c(\pm h_{\parallel})\!=\!\frac{I_c(0)}{\text{E}(e^2)} \sqrt{ 
\left[ \int_{0}^{\pi/2} \!\!\!\!\! \mathcal{I}(\tau) \sin(h_{\parallel} \sin\tau) d\tau \right]^2 \!+\!
\left[ \int_{0}^{\pi/2} \!\!\!\!\! \mathcal{I}(\tau) \cos(h_{\parallel} \sin\tau) d\tau \right]^2},
\end{equation}

\noindent where $I_c(0)=J_c WL$ is the \jun zero-field critical current and $\text{E}(e^2)= \text{E}(\pi/2,e^2)$ is the \textit{complete} elliptic integral of the second kind of real argument $e^2\leq 1$. Eq.(\ref{Icquarter}) is very general and, by properly adjusting the integration limits, applies to a junction shaped as any arc of ellipse with $\tau$ in an arbitrary interval $[\tau_1,\tau_2]$.

\subsection{Threshold curves computations}

\begin{figure}[tb]
\centering
\includegraphics[width=8cm]{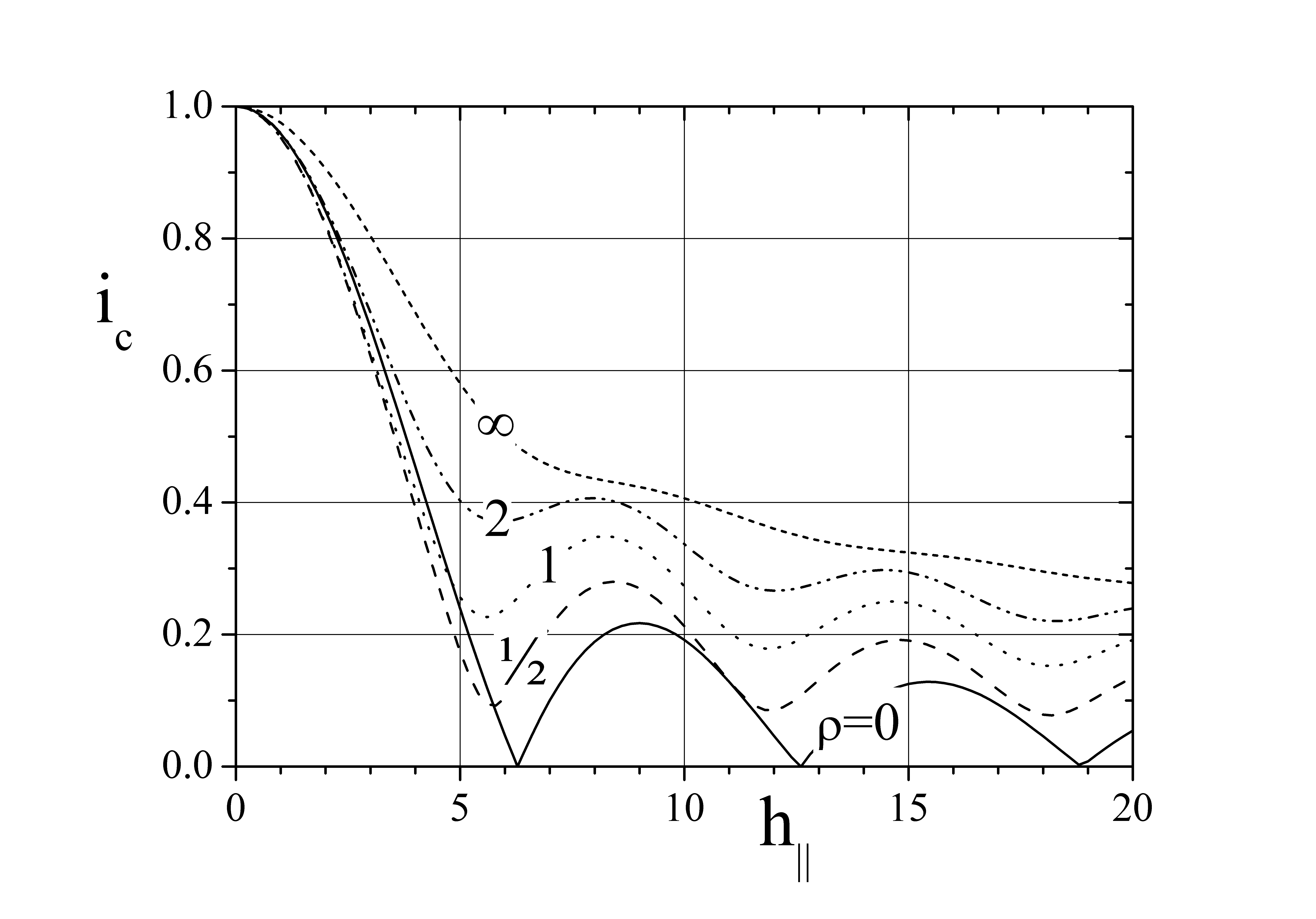}
\caption{Magnetic diffraction patterns of quarter-elliptic \juns for different values of the axes ratio, $\rho$.}
\label{MDPquarter}
\end{figure}

Unfortunately, no analytical closed form exists for the definite integrals in Eq.(\ref{Icquarter}) which means that, in general, the MDP of a quarter-\elli \jun has to be computed numerically. The theoretical magnetic dependence of the normalized critical current, $i_c(\pm h_{\parallel}) =I_c(\pm h_{\parallel})/I_c(0)$, of a small quarter-\elli \ann\jun is shown for different $\rho$ values in Figure~\ref{MDPquarter}. We observe that for quarter-circular \juns ($\rho=1$), Eq.(\ref{Icquarter}) results in a Fresnel (or near-field) MDP typical of short \Jos \juns with asymmetric boundary conditions\cite{JAP10}:
\vskip -10pt
$$ i_c(\pm h_{\parallel})= \sqrt{ \text{J}_0^2(h_{\parallel}) + \text{H}_0^2(h_{\parallel})}=\sqrt{\frac{ S^2(\sqrt{2h_{\parallel}/\pi})+ C^2(\sqrt{2h_{\parallel}/\pi}) }{2h_{\parallel}/\pi}}.$$

\noindent In the above expression $\text{J}_n$ and $\text{H}_n$ are, respectively, the $n$-th order Bessel function of the first kind and the $n$-th order Struve function.  We have also introduced the Fresnel's Sine and Cosine integrals defined, respectively, by: $S \left(\sqrt{{2h_{\parallel}}/{\pi}} \right)=\sqrt{{2h_{\parallel}/\pi}}\, \int_0^1 \sin h_{\parallel} x^2 dx$ and $C \left( \sqrt{{2h_{\parallel}}/{\pi}} \right)= \sqrt{{2h_{\parallel}/\pi}}\,\int_0^1 \cos h_{\parallel} x^2 dx$ ($h_{\parallel}>0$). In the limit $h_{\parallel} \to 0$, $S\approx 0$ and $C\approx \sqrt{2h_{\parallel}/\pi}$, so that, as required, $i_c(0)=1$. In the opposite limit, i.e., for $h_{\parallel} \to \infty$, $S\approx C \approx 1/2$ and $i_c(h_{\parallel})$ asymptotically decreases as $1/\sqrt{h_{\parallel}}$. It is also interesting to observe that for a very prolate annulus, $b\!>>\!a$, i.e., $-e^2\approx \rho^2\!>>\!1$, the normalize MDP reaches a $\rho$-independent shape:
\vskip -10pt
$$ i_c(\pm h_{\parallel})= \frac{\pi}{2}\sqrt{ \text{J}_1^2(h_{\parallel}) + \text{H}_{-1}^2(h_{\parallel})},$$
\noindent which for large fields decreases as $\sqrt{\pi/h_{\parallel}}$.

\noindent For a semi-\elli \ann \jun with $\tau\in [-\pi/2,\pi/2]$, taking into account the $\phi-\phi_0$ symmetry, Eq.(\ref{Icquarter}) reduces to:
\vskip -10pt
\begin{equation} \label{Icsemi}
i_c(\pm h_{\parallel})= \frac{1}{\text{E}(e^2)}\left| \int_{0}^{\pi/2} \!\!\!\!\!     \mathcal{I}(\tau) \cos(h_{\parallel} \sin\tau) d\tau \right|.
\end{equation}

\begin{figure}[tb]
\centering
\subfigure[ ]{\includegraphics[width=7cm]{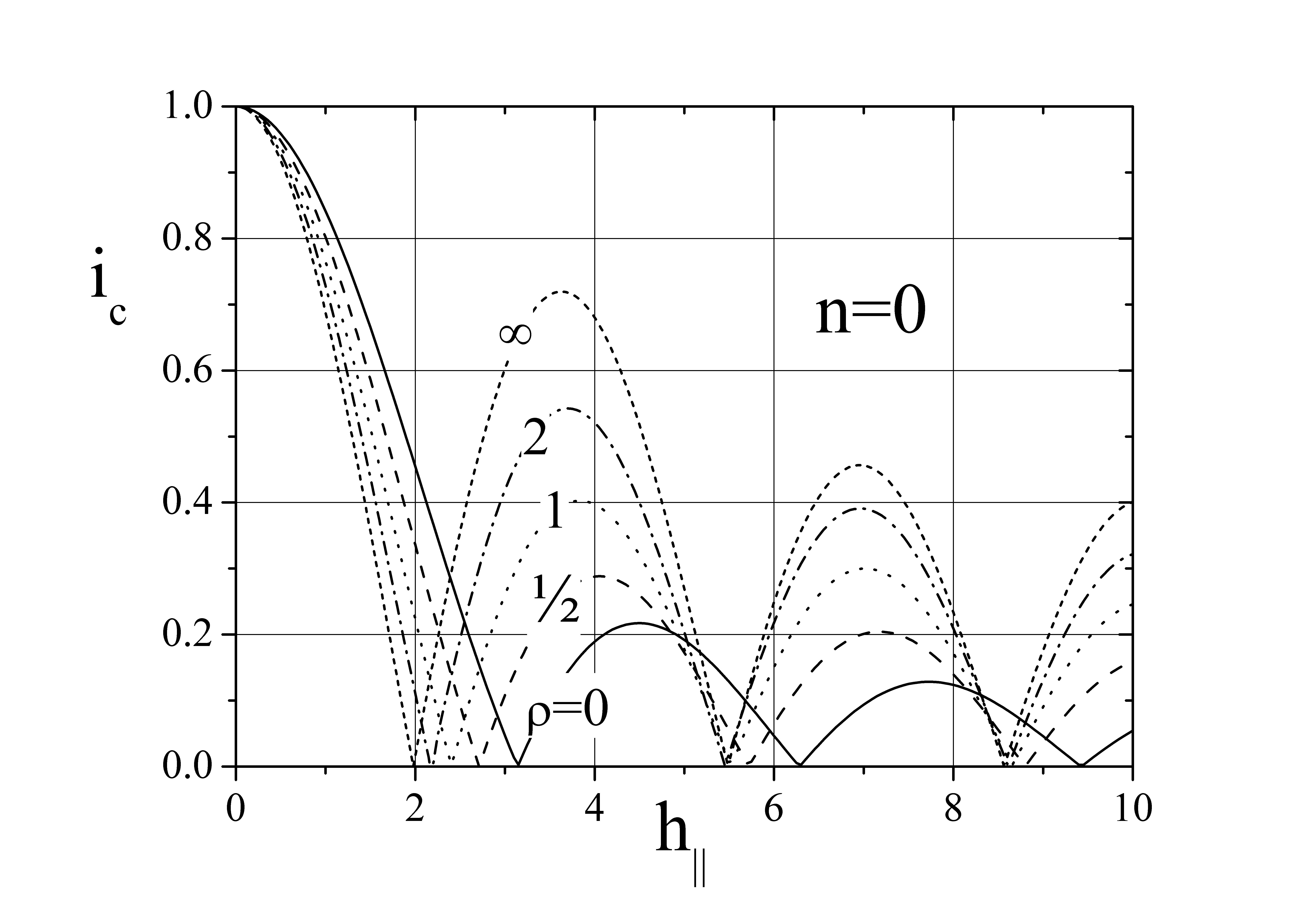}}
\subfigure[ ]{\includegraphics[width=7cm]{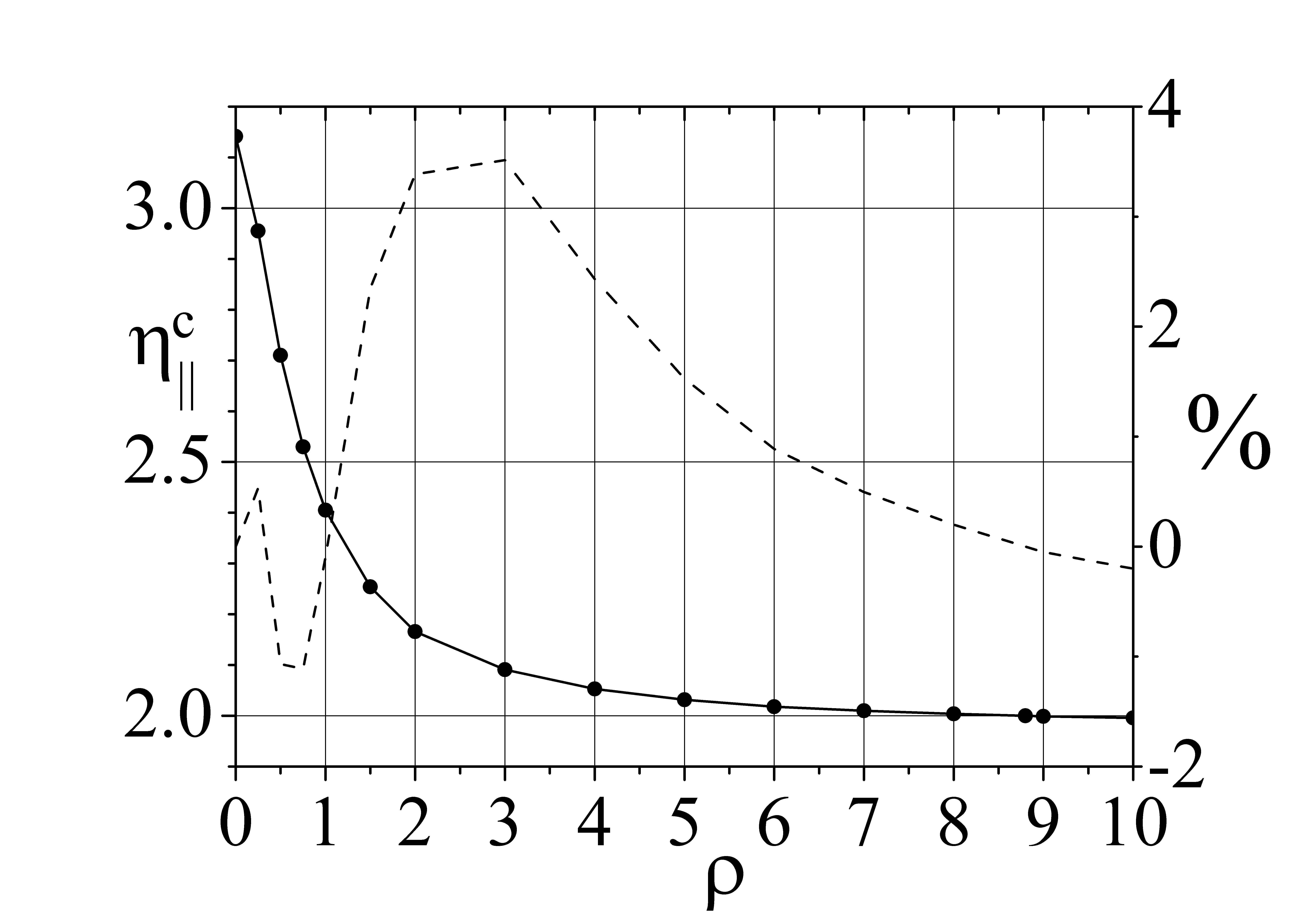}}
\caption{(a) Threshold curves of semi-elliptic and \elli \ann \juns according to Eq.(\ref{Icsemi}); n=0 indicates the absence of trapped Josephson vortices.(b) The normalized critical field $h^c_{\parallel}$ vs. $\rho$: the dots are the numerical values, while the solid line is the function $2\text{E}(\text{ArcTanh}^2\,\rho)$; the dashed line shows the relative error (right scale).}
\label{etacvsrho}
\end{figure}

\noindent This expression also applies to full ellipses for which the parameter $\tau$ spans over a $2\pi$ range (as far as we disregard trapped Josephson vortices). The MDPs in Eq.(\ref{Icsemi}) are shown in Figure~\ref{etacvsrho}(a) for different $\rho$ values. For linear \juns ($e^2=1$), we again recover the expected Fraunhofer-like MDP, $|\text{Sinc}\, h_{\parallel}|=|j_{0}(h_{\parallel})|$, where $j_0$ is the zero order spherical Bessel function of first kind. For $e^2=0$ we end up with the Bessel-like dependence $|\text{J}_0(h_{\parallel})|$ of circular and semi-circular junctions which first nulls for $h_{\parallel}\simeq 2.408 \simeq \pi/(1+\pi^{-1})$. At last, for a very prolate ellipse, $b\!>>\!a$, the normalize MDP reaches the limit $\rho$-independent shape $\pi|\text{H}_{-1}(h_{\parallel})|/2$. We observe that, as $\rho$ increases, the secondary lobes grow, i.e., the MDPs look more and more like those of a linear \jun with the \Jos current density, $J_c(\tau)$, peaked at the edges\cite{barone} ($\tau=\pm \pi/2$).

\noindent It is evident that for an \elli \jun in a uniform field the minima in the magnetic pattern are not integer multiples of the first one, although they are (almost) equally spaced, the separation between two contiguous minima being about $\pi$. It is also worth to note that while the axes ratio $\rho$ changes from $0$ to $\infty$, the first critical field, $h^c_{\parallel}$, i.e., the first zero of the MDP only changes from $\pi$ to $2$\cite{note}. This suggests that the dependence of $h^c_{\parallel}$ on $\rho$ involves the complete \elli integral of the second kind, $\text{E}(k^2)$, that spans the range $[1,\pi/2]$; Figure~\ref{etacvsrho}(b) compares the numerically found values (dots) of $h^c_{\parallel}$ with the empirical expression (solid line) $h^c_{\parallel}(\rho)=2\text{E}(\text{ArcTanh}^2\,\rho)$ together with the percentage relative difference between the numerical findings and the proposed dependence (dashed line referred to the right scale). In passing, we remark that the integrals in  Eq.(\ref{Icsemi}) can be accurately reproduced by the following empirical functions:
\vskip -10pt
\begin{equation}
i_c(\pm h_{\parallel},\rho)=
\Gamma(1+\sigma)\left(\frac{2}{|h_{\parallel}|}\right)^\sigma \left|J_\sigma (h_{\parallel})\right|,
\label{empir}
\end{equation}

\noindent where $\Gamma$ is the Gamma function of real argument and  $2\sigma (\rho)= (1+\pi^{-1})h^c_{\parallel}(\rho)-\pi$. 

\subsection{Periodic boundary conditions}

For any (real) values of $a$ and $b$, it is possible to find a value, $c$, and a number, $\nu$, such that $a=c \cosh \nu$ and $b=c \sinh \nu$; $c=\pm\sqrt{a^2-b^2}$ are the abscissae of the ellipse's foci. For a circle $a=b$, then $c=0$ and $\nu=\infty$; for a segment, $b=0$, then $c=a$ and $\nu=0$. Furthermore, when $|b|>|a|$, then the foci lie on the $Y$-axis and $\nu=\text{ArcTanh}\,\rho$ is a complex number. $(\nu,\tau)$ forms the so-called planar elliptic coordinate system. In general, the periodic conditions for the two-dimensional field $\phi$ around an EAJTJ are written as:

\vskip -10pt
\begin{subequations}
\begin{eqnarray} \label{peri1}
\phi(\nu,\tau+2\pi)=\phi(\nu,\tau)+ 2\pi n,\\
\phi_\tau(\nu,\tau+2\pi)=\phi_\tau(\nu,\tau),
\label{peri2}
\end{eqnarray}
\end{subequations}

\noindent $n$ being an integer number corresponding to the algebraic sum of \Jos vortices (or fluxons) trapped in the \jun at the time of the normal-to-superconducting transition. Eqs.(\ref{peri1}) and (\ref{peri2}) were derived in Ref.\cite{PRB96} for ring-shaped \juns and state that observable quantities such as the \Jos current (through $\sin\phi$) and the magnetic field (through $\phi_\tau$) must be single valued upon a round trip. The net number of trapped fluxons comes out to be the algebraic difference between the number of flux quanta associated with the fluxoids in each electrode\cite{vanDuzer}. $\phi$ in Eq.(\ref{phiditau})	is $2\pi$-periodic, therefore, to implement the periodic conditions, one only has to add a term, $n\tau$, that accounts for $n$ distributed $2\pi$-kinks:
\vskip -15pt
\begin{equation}
\label{phiditau1}
\phi(\tau) = h_{\parallel} \sin \, \tau + n\tau+ \phi_0,
\end{equation} 

\noindent Being $\phi$ still an odd function (disregarding $\phi_0$), we can use again Eq.(\ref{Icquarter}) to derive the most general expression for the threshold curve of a short EAJTJ:	
\vskip -10pt
\begin{equation} \label{Icelli}
i_c(\pm h_{\parallel},n)= \frac{1}{2\text{E}(e^2)}\left| \int_{0}^{\pi}  \mathcal{I}(\tau) \cos(h_{\parallel} \sin\tau + n\tau) d\tau \right|.
\end{equation}

\begin{figure}[tb]
\centering
\includegraphics[width=7cm]{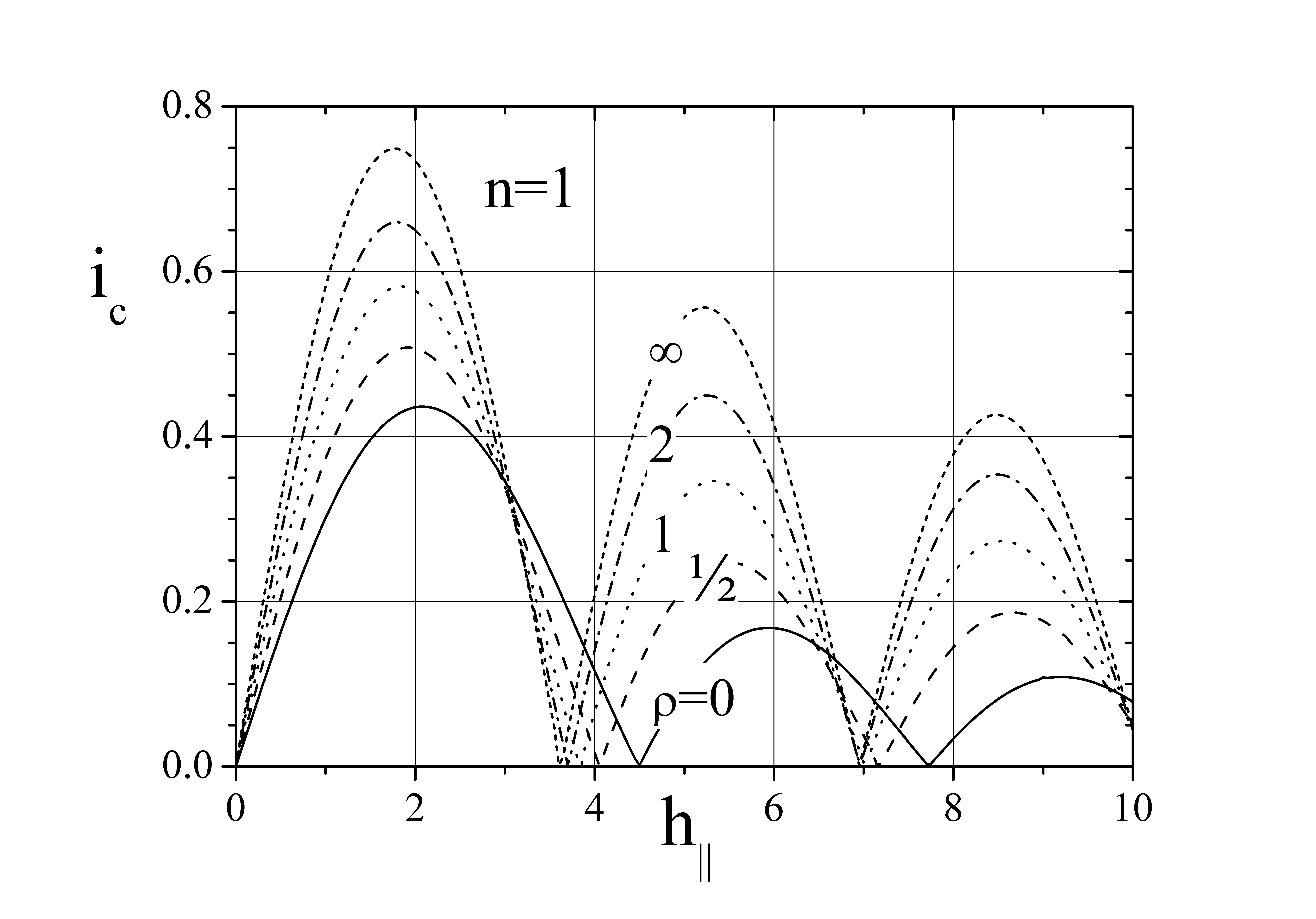}
\caption{Magnetic diffraction patterns of \elli\ann\juns as given by Eq.(\ref{Icelli}) with $n=1$ for different values of the axes ratio, $\rho$ (compare to Figure~\ref{etacvsrho}(a) for $n=0$).}
\label{MDP1}
\end{figure}

\noindent The MDPs of Eq.(\ref{Icelli}) with $n=1$ are plotted in Figure~\ref{MDP1} for different values of $\rho$. As we progressively decrease the ellipse minor semi-axis $b$ (while keeping the major semi-axis $a$ constant) an EAJTJ tends to a $2W\times 2a$ rectangular junction with a slit in the middle; with $n=e^2=1$, Eq.(\ref{Icelli}) reduces to: $i_c(\pm h_{\parallel},1)= \left| \frac{\sin h_{\parallel} /h_{\parallel} - \cos h_{\parallel}}{h_{\parallel}} \right|=|j_1(h_{\parallel})|$, where $j_1$ is the order one spherical Bessel function of first kind. In general, if $n$ Josephson vortices are trapped in a slit realized in a linear junction, then $i_c(\pm h_{\parallel},n)=|j_n(h_{\parallel})|$, in analogy with ring-shaped \juns for which $i_c(\pm h_{\parallel},n)=|J_n(h_{\parallel})|$. Ultimately,   generalizing Eq.(\ref{empir}), we have proved that the integrals in Eq.(\ref{Icelli}) can be approximated by normalized functions which involve the Bessel function of nonintegral order:
\vskip -10pt
$$i_c(\pm h_{\parallel},\rho,n)= \Gamma(1+n+\sigma)\left(\frac{2}{|h_{\parallel}|}\right)^\sigma \left| J_{n+\sigma} (h_{\parallel}) \right|.$$

\noindent To correctly interpret Figures~\ref{MDPquarter}, \ref{etacvsrho}(b) and \ref{MDP1}, one should keep the semi-axis $a$ constant and change $\rho$ through the semi-axis $b$. If $a$ changes, then the field normalization changes.

\section{Long \elli \ann junctions} 

In this section we derive the appropriate partial differential equation (PDE) for an EAJTJ in an external magnetic field. The total tunnel current density is given by:
\vskip -10pt
$$J_z=J_c\sin \phi +\frac {\Phi_0}{2\pi R}\phi_t,$$

\noindent where the last term takes into account the quasi-particle tunnel current assumed to be ohmic, i.e., $R$ is the voltage independent quasi-particle resistance per unit area. The subscripts on $\phi$ denote partial derivatives. Following Refs.\cite{gronbech,goldobin01}, a one-dimensional planar \textit{curved} \Jos tunnel \jun of constant width in the presence of a barrier-parallel external magnetic field, ${\bf {H^e}}$, is described by the following partial differential equation for $\phi$:
\vskip -10pt
$$\lambda _j^2 \frac{d^2 \phi}{ds^2} - \frac 1{\omega_{p}^2}\phi _{tt}-\sin \phi =\gamma(s)+ \frac{\phi _t}{\omega_p^2 R c_s} -\lambda_J^2\epsilon_r\epsilon_0 R_s \frac{d^2 \phi_t}{ds^2} + \frac{\Delta}{J_c}  \frac{d {\bf {H^e}} \cdot {\bf \hat{\nu}}}{ds}$$


\noindent where $\omega_{p}^2={2\pi J_c}/{\Phi_0 c_s}$, and $c_s$ being the specific junction capacitance. $\gamma(s)$ is the local normalized bias current density and $\Delta$ is the coupling between the external field and the field in the junction\cite{gronbech,PRB96}. Here $s$ is a curvilinear coordinate along the junction. It is well known that $\lambda_J$ gives a measure of the distance over which significant spatial variations of the phase occur; the plasma frequency, $\omega_p/2\pi$, is the oscillation frequency of small amplitude waves. ${\bf\hat{\nu}}\equiv (x_0/a^2N,y_0/b^2N)$ is the (outward) normal unit vector to the ellipse $x^2/a^2+y^2/b^2=1$ at the point $(x_0=a\sin\tau,y_0=b\cos\tau)$ with  $N=\sqrt{x_0^2/a^4+y_0^2/b^4}$. Further, we can introduce the parameter $\overline{c}=\omega _{p}\lambda _j$ which gives the velocity of electromagnetic waves in the \jun and is called the Swihart velocity. The third term in the right-hand side takes into account the effect of the surface currents in the London layers, i.e., $R_s$ is the voltage independent surface resistance per area. Introducing the dimensionless loss coefficients $\alpha^{-1} ={\omega _{p} R c_s}$ and $ \beta =\epsilon _r\epsilon _0\omega _{p}R_s$  and normalizing the time $t$ to $\omega _{p}^{-1}$, the last equation takes the form:
\vskip -10pt
\begin{equation}
\label{psge}
\lambda _j^2 \frac{d^2 \phi}{ds^2} - \phi _{tt}-\sin \phi =\gamma(s) + \alpha \phi_t-\lambda_J^2 \beta \frac{d^2 \phi_t}{ds^2}+ \frac{\Delta}{J_c} \frac{d H_\nu}{ds},
\end{equation}


\noindent where $H_\nu= {\bf {H^e}} \cdot {\bf \hat{\nu}}$ is the component of the externally applied in-plane magnetic field normal to the \jun perimeter. Therefore, in the experiments the magnetic field can be substituted by a properly chosen additional bias current $\gamma_1(s)$ and vice versa. Eq.(\ref{psge}) is called Perturbed sine-Gordon Equation (PSGE). Because of its local form, it is quite general and holds for \juns of any geometrical shape. We note that the first and last terms on the right-hand side of Eq.(\ref{psge}), from a mathematical point of view, play the same role. 


\subsection{The sine-Gordon modeling} 

We now want first to compute the normal component, $H_\nu$, of a uniform magnetic field ${\bf {H}}\equiv(0,H_\parallel)$. $H_\nu=H_y \hat{\nu}_y=H_{\parallel} \mathcal{I}^{-1}(\tau) \cos\tau$ is maximum at the ellipse poles, $H_\nu(0)=-H_\nu(\pi)=H_{\parallel}$, and vanishes at the ellipse equatorial points, $\tau=\pm \pi/2$. Next, recalling that $\frac{d }{ds}= \frac{d\tau}{ds} \frac{d }{d \tau}= \frac{1}{a \mathcal{I}}\frac{d }{d \tau}$, the directional derivative of the normal field is:
\vskip -10pt
\begin{equation}
\label{dirder}
\frac{d H_{\nu}}{ds}=\frac{H_\parallel  \rho^2 }{a \mathcal{I}^{4}}  \sin\tau.
\end{equation}

\noindent It can be shown that the second directional derivative is:
\vskip -10pt
\begin{equation}
\label{laplacian} 
\frac{d^2}{ds^2} = \frac{1}{a^2 \mathcal{I}^2} \left[ \frac{d^2}{d\tau^2} + \frac{(1-\rho^2)}{\mathcal{I}^2} \sin\tau \cos\tau \frac{d}{d\tau}\right].
\end{equation}

\begin{figure}[tb]
\centering
\includegraphics[width=7cm]{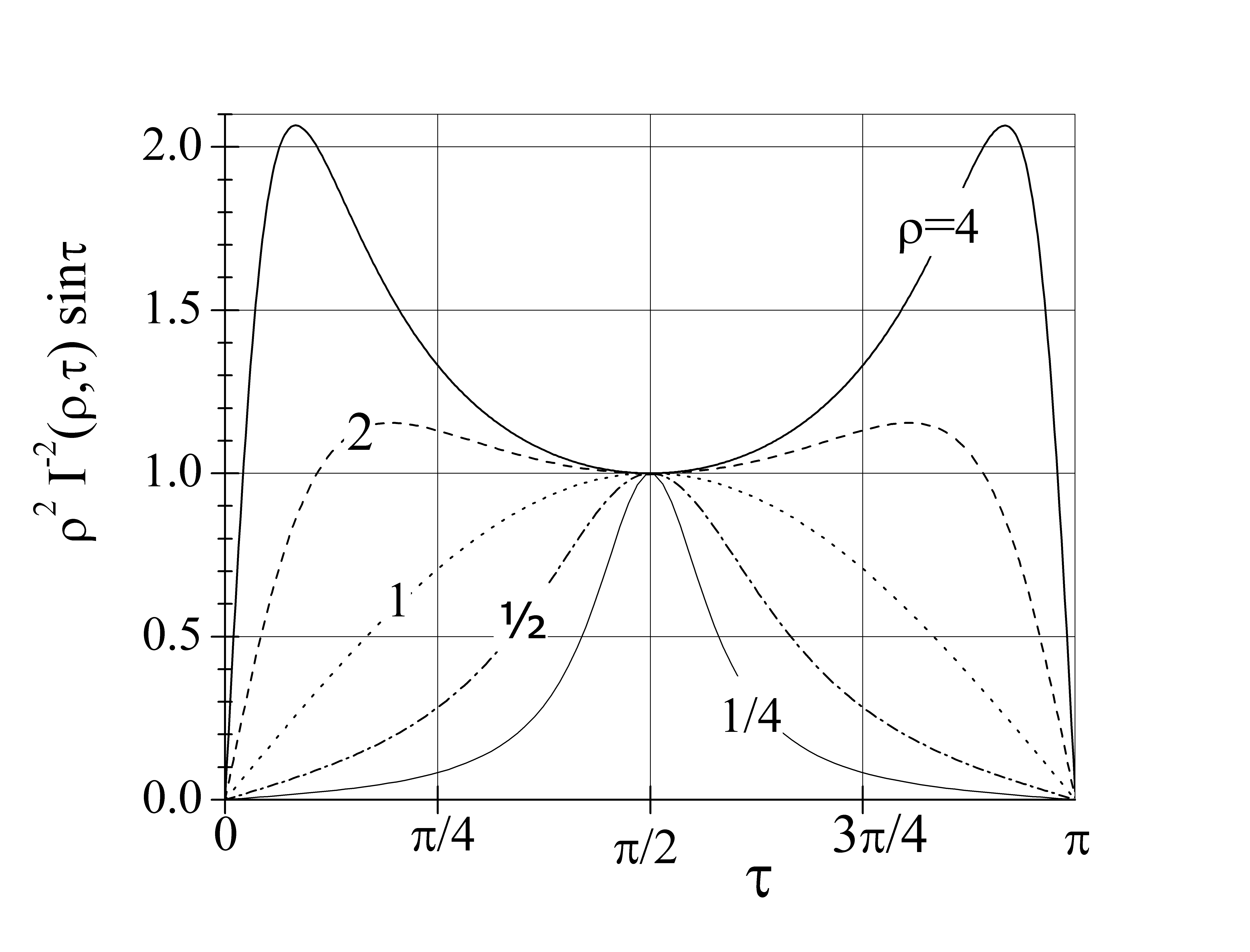}
\caption{Plot of $\rho^2 \,\mathcal{I}^{-2}(\rho,\tau) \sin\tau$ for $0\leq\tau \leq \pi$ and different $\rho$ values.}
\label{RhoOverIsquared}
\end{figure}

\noindent Inserting Eqs.(\ref{dirder}) and	(\ref{laplacian}) in Eq.(\ref{psge}), we end up with the following PSGE for an EAJTJ:
\vskip -10pt
$$\phi_{\tau\tau} + \frac{1-\rho^2}{\mathcal{I}^2(\rho,\tau)} \sin\tau \cos\tau\, \phi_\tau + \beta \left( \phi_{\tau\tau t} + \frac{1-\rho^2}{\mathcal{I}^2} \sin\tau \cos\tau\, \phi_{\tau t}\right) = $$
\vskip -10pt
\begin{equation}
\label{diff1}
= \left[\frac{a\,\mathcal{I}(\rho,\tau)}{\lambda_J}\right]^2[\phi_{tt}+ \sin \phi +\gamma(\tau)  + \alpha \phi_t ] +   h_{\parallel}\Delta \frac{\rho^2}{\mathcal{I}^2(\rho,\tau)} \sin \tau,
\end{equation}
\noindent where again $h_{\parallel}=\kappa a H_{\parallel}$. Eq.(\ref{diff1}) states that for an \elli \ann \jun the magnetic field enters directly into the PSGE in contrast to the case of linear junctions for which it appears only in the boundary conditions. Further, the different sections of the annulus {\it feel} different fields; diametrically opposed points {\it feel} opposite fields and the field term in Eq.(\ref{diff1}) is out of phase with respect to the actual normal field. Moreover, the effect of a given field depends quadratically on the axes ratio $\rho$; the odd expression $\rho^2\,\mathcal{I}^{-2} \sin \tau$, whose rms value is $\sqrt{\rho/2}$, is plotted in Figure~\ref{RhoOverIsquared} for different $\rho$ values.
\noindent For $\rho=\mathcal{I}^2(\rho,\tau)=1$, Eq.(\ref{diff1}) reduces to well studied PSGE for ring-shaped junctions\cite{gronbech,PRB96}. It is also worth to notice that, in the limit $\rho \to 0$, Eq.(\ref{diff1}) reproduces the classical PDE for a linear junction. In fact, observing that $\mathcal{I}^2(0,\tau)= \cos^2\tau$ and that, as $\rho$ gets smaller and smaller, $\rho^2 \mathcal{I}^{-2}(\tau) \sin \tau$ approaches the unitary impulse function, $\delta_1(\pm \pi/2)=\pm 1$ and zero elsewhere (see Figure~\ref{RhoOverIsquared}), we have:
\vskip -15pt
$$\phi_{\tau\tau} +  \tan\tau\, \phi_\tau =
\left(\frac{a}{\lambda_J}\right)^2 \cos^2\!\tau\left( \phi_{tt}+ \sin \phi + \gamma  + \alpha \phi_t \right) - \beta \left( \phi_{\tau\tau t} + \tan\tau\, \phi_{\tau t} \right)+ h_{\parallel}\Delta \delta_1(\tau).$$

\noindent Now recalling that $x=a\sin\tau$, it is not difficult to derive that for any composite function $\phi(x(\tau))$ it is $\phi_{\tau\tau} +  \tan\tau\, \phi_\tau=a^2\cos^2\tau\, \phi_{xx} $; introducing the normalized spatial coordinate $\xi=x/\lambda_J$, we recognize the well known PSGE for a linear overlap JTJ:
\vskip -10pt
$$\phi_{\xi\xi} - \phi_{tt}- \sin \phi = \gamma  + \alpha \phi_t - \beta  \phi_{\xi\xi t} + h_{\parallel}\Delta \delta_1(a/\lambda_J),$$

\noindent with boundary conditions $\phi_\xi(a/\lambda_J)= -\phi_\xi(-a/\lambda_J)\propto h_{\parallel}$. In concluding, Eq.(\ref{diff1}) is more general that it appears at a first glance.

\subsection{The static numerical simulations} 

In what follows, we are interested in the static, i.e., time-independent solutions of Eq.(\ref{diff1}) and, in order to have zero fluxons trapped in the junction, the conditions on the phase periodicity are: $ \phi(\tau+2\pi)=\phi(\tau)$ and $\phi_\tau(\tau+2\pi)=\phi_\tau(\tau)$. The direct numerical integration of Eq.(\ref{diff1}) with $\alpha=\beta=0$ poses large problems of stability due to the fact that there is no loss in the system\cite{PRB10}; to avoid this problems, we set $\alpha=3$ in order to have a fast decay towards a static solution (in real \juns $\alpha \le 0.1$). The term containing the surface loss was simply dropped to save computer time ($\beta=0$). For the sake of simplicity, in our theoretical investigation the bias current was supposed to be uniform, $\gamma(\tau)=\gamma$, and $\Delta$ was set to 1; in real devices, depending on the specific electrodes geometry, non uniform bias distribution\cite{vasenko}, self-field\cite{yama,SUST13a} and focusing\cite{vanDuzer,lee} effects should be taken into account. In addition, we introduced the field normalization usual for long linear JTJs, $h'_{\parallel}=H_{\parallel}/J_c \lambda_J=h_{\parallel} \lambda_J d_j/a d_m$; quite often in the literature the factor $d_j/d_m \geq 1$ has been ignored\cite{note2}. With symmetric bias and field terms, in the steady state regime with $n=0$, the \Jos phase has the symmetry $\phi(\pm \pi/2+\tau)=\phi(\pm \pi/2-\tau)$, therefore Eq.(\ref{diff1} can be integrated in the range $[-\pi/2,\pi/2]$ with boundary conditions $\phi_\tau(\pm \pi/2)=0$. The commercial finite element simulation package COMSOL MULTIPHYSICS (www.comsol.com) was used to numerically solve Eq.(\ref{diff1}) for different values of the normalized semi-axis, $a/\lambda_J$, and of the mean axes ratio, $\rho$. Specifically, we have numerically computed the maximum value, $i_c$, of the zero-voltage bias vs. $h'_{\parallel}$. To begin with, we have checked that for $a<0.1 \lambda_J$ we were able to accurately reproduce all plots in Figure~\ref{etacvsrho}(a). 

\begin{figure}[tb]
\centering
\subfigure[ ]{\includegraphics[width=7cm]{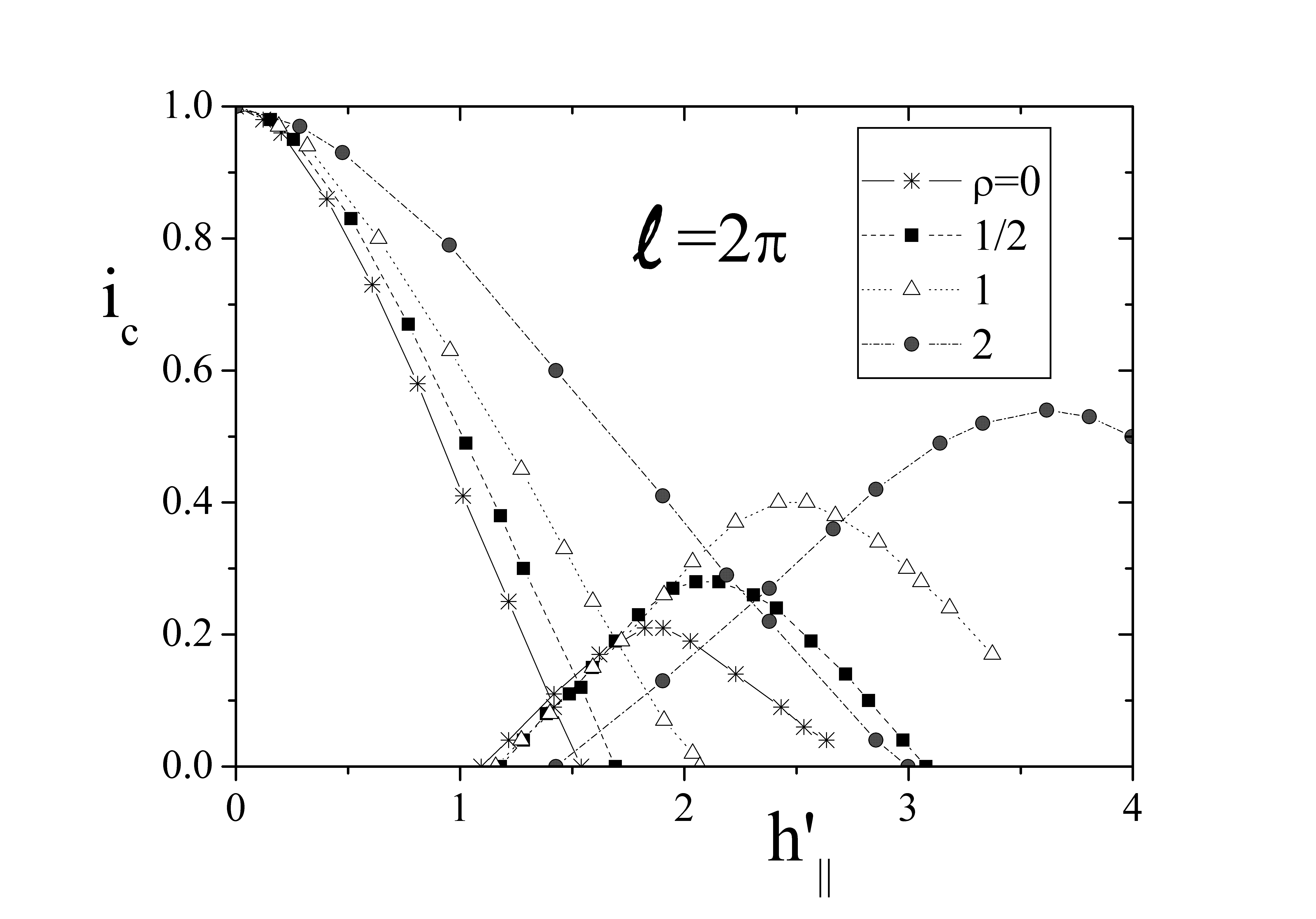}}
\subfigure[ ]{\includegraphics[width=7cm]{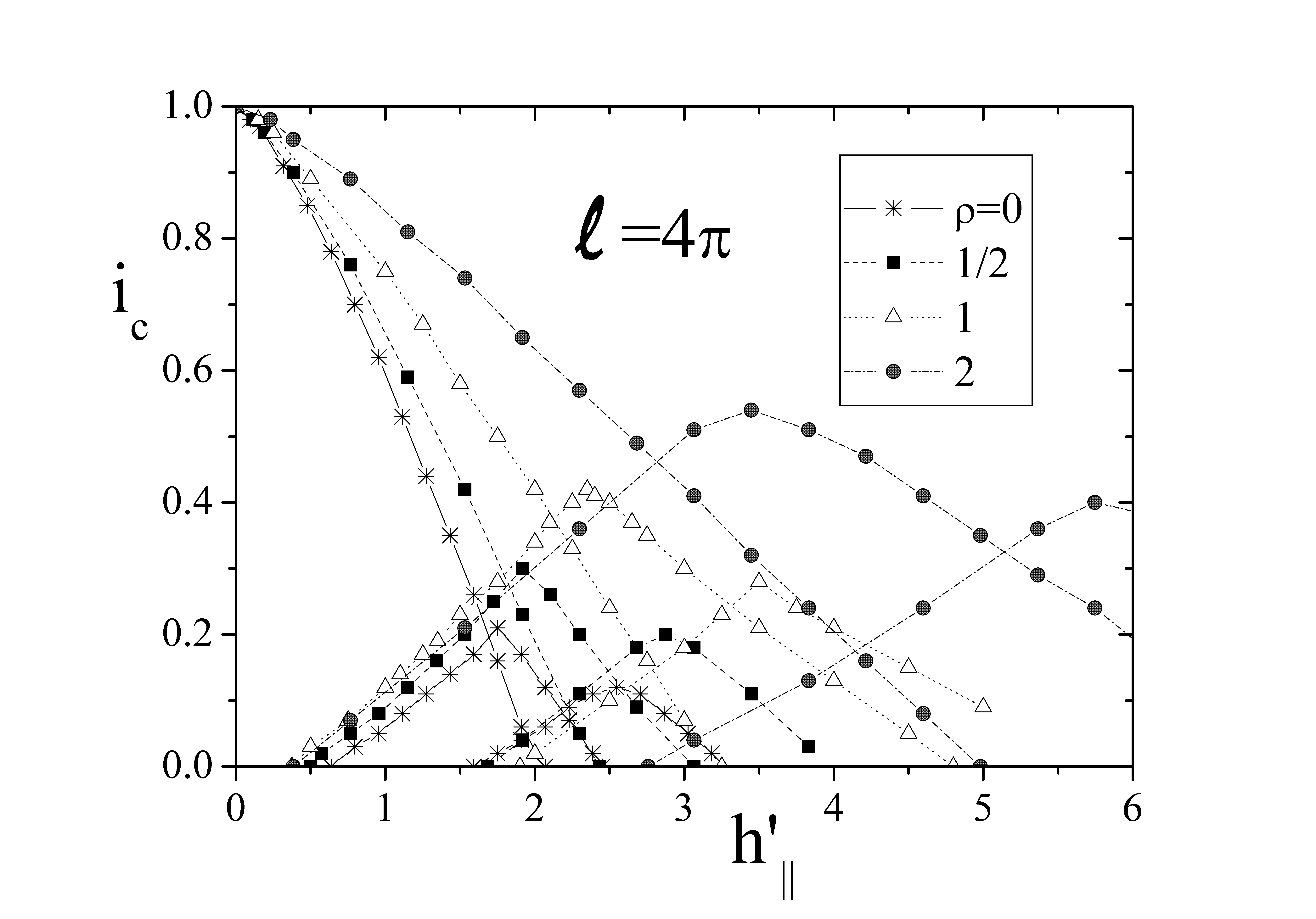}}
\caption{Numerical magnetic diffraction patterns of long EAJTJs for different values of the axis ratio, $\rho$, and two normalized perimeters: (a) $\ell=2\pi$ and (b) $\ell=4\pi$. The magnetic field is normalized to $J_c\lambda_J$.}
\label{2and4Pi}
\end{figure}

\noindent For long EAJTJs we wish to confront the MDPs of samples having the same normalized perimeter, $\ell = P/\lambda_J=4aE(e^2)/\lambda_J$, and different axes ratios; accordingly, for given $\ell$ and $\rho$, we must set $a/\lambda_J=\ell/4E(1\!-\!\rho^2)$. In order to compare the numerical findings with the experimental data, we will limit our interest to $\rho=0.5$ [$\text{E}(0.75)\approx 1.21$] and $\rho=2$ [$\text{E}(-3)\approx 2.42$], while the cases of well understood linear geometry, i.e., $\rho=0$ [$\text{E}(1)=1$] and circular geometry, i.e., $\rho=1$ [$\text{E}(0)=\pi/2$], are taken for reference. Pronounced deviations from the theoretical behavior of short \juns were found for $\ell=2\pi$, as can be seen in Figure~\ref{2and4Pi}(a) where the $i_c(h'_{\parallel})$ dependence is reported. For small fields a long linear JTJ behaves as a perfect diamagnet by establishing circulating screening currents which maintain the interior field at zero. This ''Meissner'' effect is not anymore possible in long curved \juns where the magnetic field does penetrate the barrier even when small; nevertheless, still the critical current decreases linearly with the applied field. We also observe that some ranges of magnetic field develop in correspondence of the pattern minima in which $i_c$ may assume two different values. In a fashion which closely resembles the behavior of long linear junctions, these values correspond to different configurations of the phase inside the barrier. In fact, each lobe in Figures~\ref{2and4Pi} is associated with a given vortex structure; at the very end of the first lobe a fluxon-antifluxon pair is present along the junction, the fluxon facing the antifluxon on diametrically opposed potential wells created by the magnetic field at the ellipse's poles. In the successive lobes the magnetic field penetrates in the barrier and more pairs are nucleated at the ellipse's equatorial points, in a way which closely recalls the behavior of the type II superconductors, even though the vortices we are dealing with are quite different from the Abrikosov vortices having a normal core. In the second lobe, for example, we start from a phase configuration very similar to that at the right side of the first lobe in which one fluxon-antifluxon pair is present in the barrier and we end up with two fluxon-antifluxon pairs, the two bunched fluxons facing the two bunched antifluxons on diametrically opposed potential wells. In order to trace the different lobes of $i_c$ vs $h'_{\parallel}$, it is crucial to start the numerical integration with a proper initial phase profile. As $l$ is increased, each higher lobe broadens further and more lobes overlap. This behavior is shown for $\ell=4\pi$ in Figure~\ref{2and4Pi}(b) where also the third lobe was computed. 

\noindent Let us recall that in Figures~\ref{2and4Pi} $i_c$ is plotted versus $h'_{\parallel}$ (rather than $h_{\parallel}$) which corresponds to a normalization of the magnetic field to $J_c\lambda_J$, as it is usual for long linear \Jos \juns for which, as $\ell$ increases the critical field saturates\cite{dettmann,franz, barone,vanDuzer,PRB96} at $2J_c\lambda_J$. For one-dimensional ring-shaped \juns it was found\cite{br96,JAP07} that the critical field is proportional to the ring radius, $r$, as far as $r\!>>\!\lambda_J$. Our numerical simulations allowed us to reach the conclusion that for long \elli annuli the critical field increases as $(a/\lambda_j)^\rho$. 

\subsection{The self-field effects}

\begin{figure}[tb]
\centering
\subfigure[ ]{\includegraphics[width=7cm]{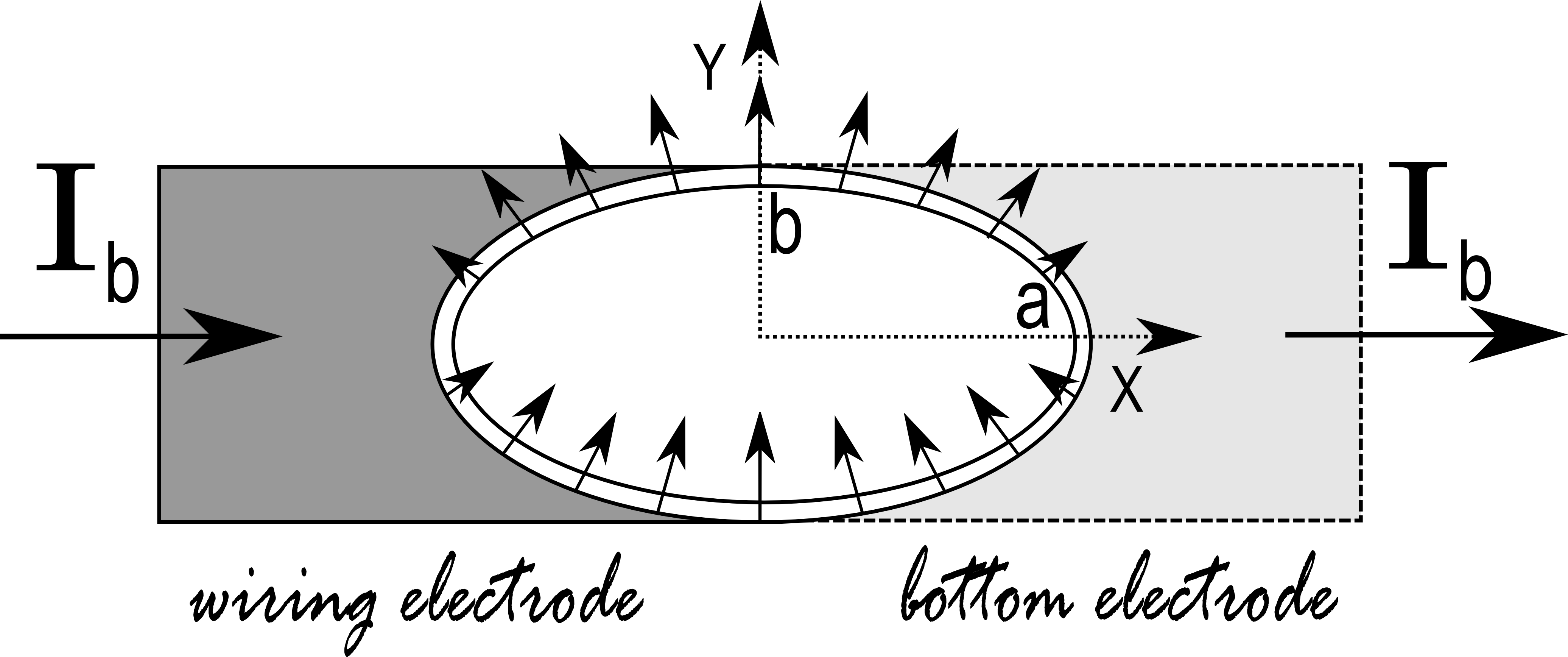}}
\subfigure[ ]{\includegraphics[width=7cm]{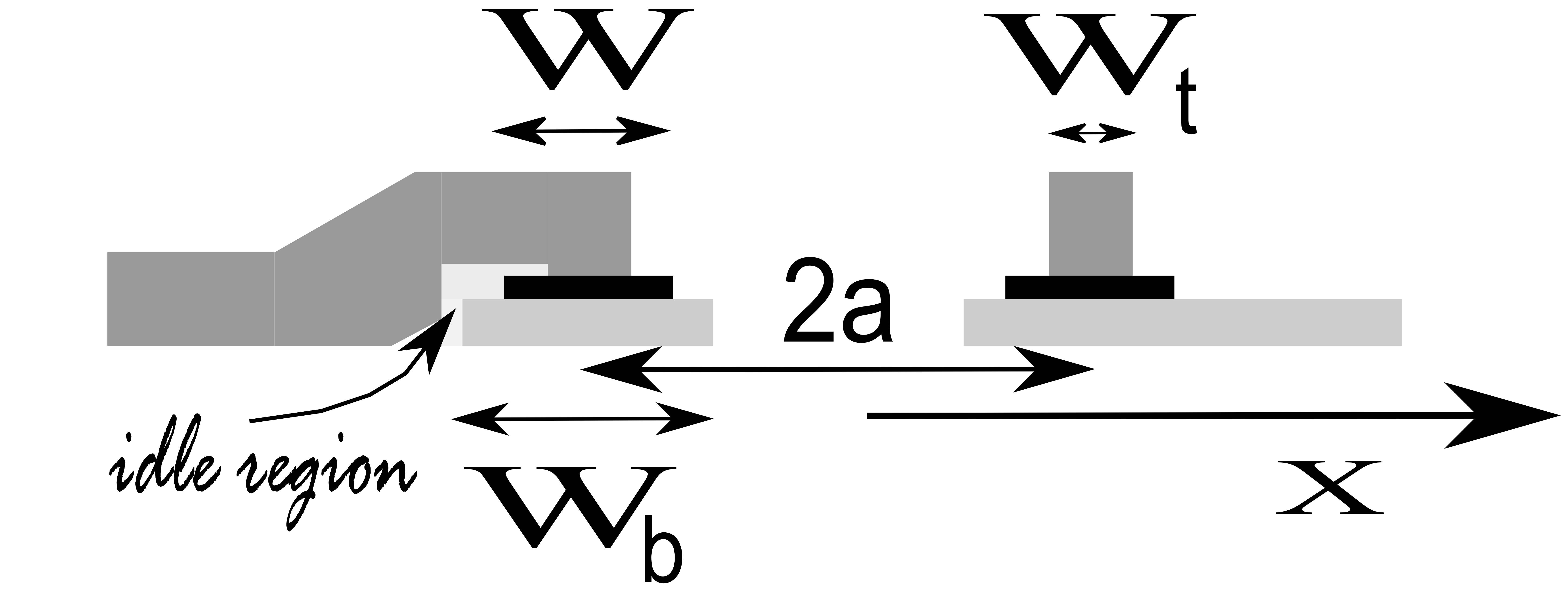}}
\caption{(a) Schematic representation of the in-plane normal magnetic self-field generated by the tangential supercurrent, $I_t$, flowing in the top electrode of a biased Lyngby-type EAJTJ with $H=0$. (b) Vertical cross section (not to scale) of our Lyngby-type \elli \ann \Jos junction.}
\label{selffield}
\end{figure}

The analysis of long EAJTJs would not be satisfactory if we neglected the effects of the magnetic field generated by the \jun feed current; for the sake of generality, we will treat them only on a qualitative basis, since this effects drastically depend on geometrical details. It will turn out that the following considerations are very useful to interpret the experimental finding reported in Section V. Let us consider an EAJTJ fed by a d.c. bias current, $I$, smaller that the critical current, $I_c$, so that it is in the stationary (zero voltage) state. As depicted in Figure~\ref{selffield}(a), $I$ is applied in the positive $X$-direction: it enters the junction, say, from the wiring electrode on the left side, gradually splits in the two arms of each electrode, recombines on the \jun right side and leaves through the bottom electrode (on the right). It is convenient to focus on one electrode at a time and we consider first the top/wiring electrode. In absence of an external magnetic field, the symmetry of the system requires that the supercurrent, $I_t(\tau)$, flowing along the top electrode within a distance of $\Lambda_t$ from the \Jos barrier is null at the equatorial points and equally distributes in each annulus branch; this supercurrent increases as we move towards the poles where it must be equal to $I/4$, since $I=I_b(0)+I_t(0)+I_b(\pi)+ I_t(\pi)$ and $I_b(0)=I_t(0)=I_b(\pi)= I_t(\pi)$. It is important to stress that $I_t$ flows within a width, $W_{t}$, that is constant all along the ellipse perimeter.
This can be better understood looking at Figure~\ref{selffield}(b) which sketches the vertical cross section along the equatorial line of our Lyngby-type EAJTJs. The oblique arrows in Figure~\ref{selffield}(a) indicate the strength and direction of the normal component of magnetic field associated (and proportional) to the tangential supercurrent $I_t$ that is inversely proportional to $W_{t}$. We can make a similar reasoning for the current, $I_b$, flowing in the bottom electrode (within a depth $\Lambda_b$) with the caveats that $I_b$ flows on the opposite side of the barrier so its normal field has opposite direction and, above all, $I_b$ flows within a width equal to (and where possible larger than) $W_{b}$ that, in turn, is larger than $W_{t}$. It means that, on the average, the normal field induced by $I_b$ is much smaller that the one generated by $I_t$ (this would still be true even if $W_{t}=W_{b}$). In our samples $W_{t}=0.3 W_{b}$; therefore, for a qualitative understanding, we will only consider the self-field contribution due to $I_t$. From Figure~\ref{selffield}(a) it is evident that such normal field has a net $Y$-component, while, in the average, the $X$-component is null. Further, the $Y$-component of the normal self-field is larger for oblate ellipses ($\rho<1$) and vice versa. The $Y$-component affects the profile of the \Jos phase, therefore, as $I$ is increased we will reach a value that makes the \jun to prematurely switch to the voltage state (the largest critical current always corresponds to the uniform phase distribution, $\phi(\tau)=\sin^{-1} I/I_c$). In the presence of an externally applied magnetic field, the system symmetry is broken and, in general, the bias current splits in unequal parts giving rise also to a net $X$-component of the normal self-field which is larger for oblate geometries ($\rho>1$). In a first approximation, as already heuristically suggested for ring-shaped junctions\cite{PRB96}, the self-field effects in long EAJTJs can be taken into account in Eq.(\ref{diff1}) by adding extra terms which simulate two uniform fields, one parallel, $\propto \rho \gamma \cos \tau$, and the other orthogonal, $\propto \rho^{-1} \gamma \sin \tau$, to the bias current flow.

\noindent We should also add that, even neglecting the self-field effects, a bias current density which is uniform in $\tau$ can never be achieved in AJTJs. If the bias current, $I$, were uniformly distributed along the height of the current carrying electrode, $\gamma(s(\tau))$ would be peaked at the equatorial points of the ellipse. However, the current in a superconducting flat film mainly flows at its boundaries\cite{vanDuzer}, therefore, more realistically, the $\gamma$ profile is depressed at the equatorial points and is largest at the poles. Although symmetric, a non-uniform current density profile, together with the self-fields, reduces the largest possible value of the critical current. In the experiments, the ratio of the zero-field critical current, $I_{c,0}$, to the current jump, $\Delta I_g$, at the gap voltage is a direct measure of this nonuniformity: the lower this ratio, the larger is the nonuniformity. 

\section{Transverse magnetic field $H_{\perp}$} 

\begin{figure}[tb]
\centering
\subfigure[ ]{\includegraphics[height=5cm]{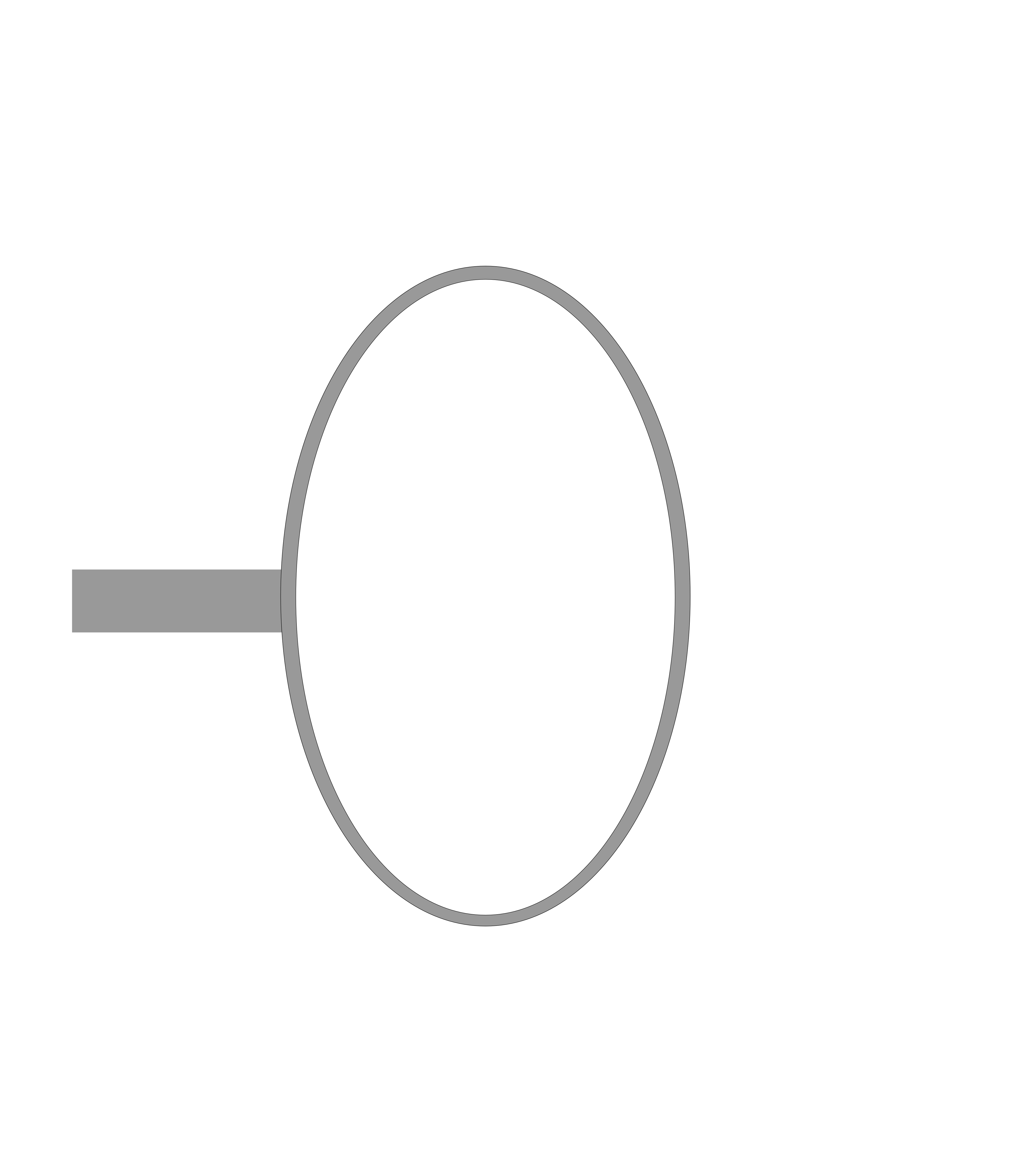}}
\subfigure[ ]{\includegraphics[height=5cm]{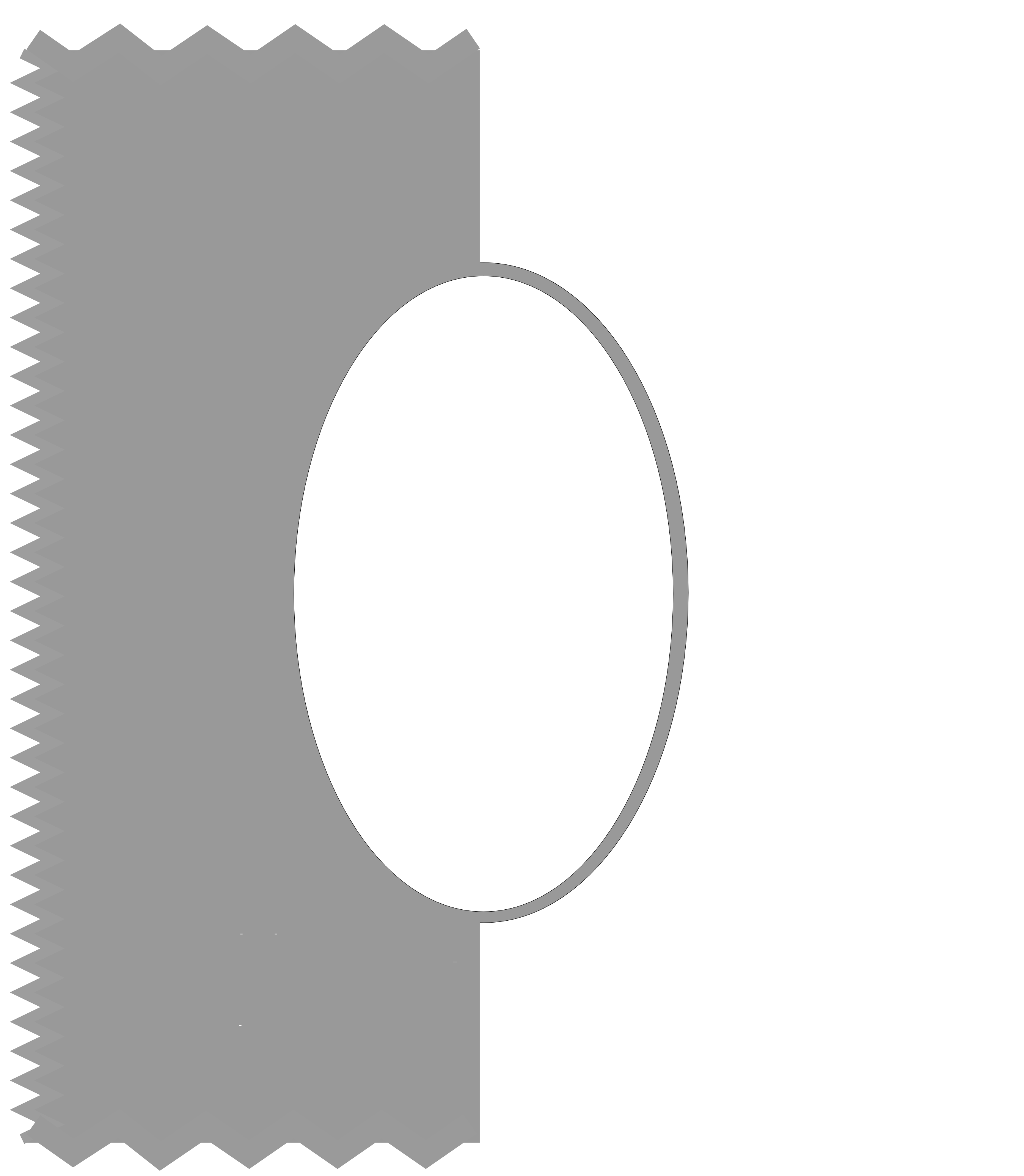}}
\subfigure[ ]{\includegraphics[height=5cm]{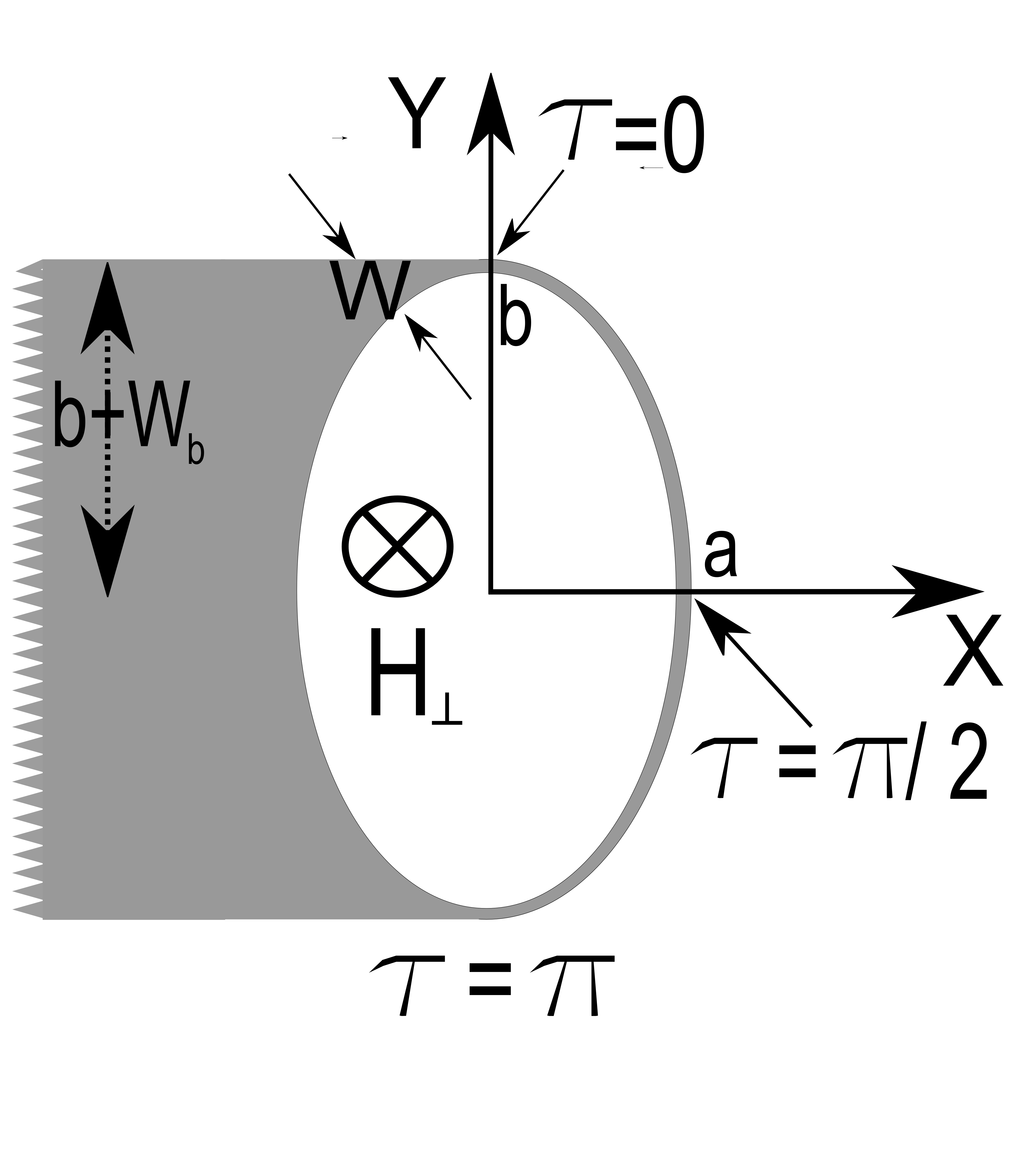}}
\caption{Sketches of the bottom electrode of \elli \ann \juns having different geometries: (a) island geometry, (b) modified Lyngby-type geometry and (c) Lyngby-type geometry. The top/wiring electrodes are specularly symmetric.}
\label{bottomelectrodes}
\end{figure}

An alternative way to modulate the critical current of a planar \Jos tunnel \jun is to apply a magnetic field, ${\bf H_{\perp}} \equiv (0,0,H_{\perp})$, perpendicular to the \jun plane\cite{rc,hf,miller}, which induces shielding currents in its electrodes. In turn, the shielding currents generate a local magnetic field whose normal component thread the \Jos barrier. The modulation amplitude drastically depends on the geometry of the electrodes and on how close to the barrier the shielding currents circulate. For rectangular \juns these effects have been already investigated both theoretically and experimentally\cite{JAP08,PRB09}; for ring shaped \juns only magnetostatic simulations and experimental data exist, and a theoretical understanding is still lacking. Here, in addition, we will analyse how a transverse field acts on an EAJTJ; more specifically, we want to derive the normal component of the magnetic field induced by the the circulating shielding currents. We will consider three different geometries formed by specularly symmetric electrodes whose bottom electrodes are shown in Figures~\ref{bottomelectrodes}(a)-(c). We will take in consideration that for window-type \juns the \elli loops in the bottom and top/wiring electrodes have the same mean axes, i.e., the same perimeter and ellipticity, but different widths (respectively, $W_b$ and $W_t$). For the sake of simplicity, the analysis will be carried out assuming that no flux is trapped in the electrode loops.

\subsection{Island geometry}
The island or $\delta$-biased\cite{PRB10} geometry, in which each electrode is fed by a very tin thread, is the easiest to analyze since in the Josephson region the electrodes have constant widths. Let us consider first the bottom electrode sketched in  Figure~\ref{bottomelectrodes}(a). When a magnetic field, $H_{\bot}$, is applied perpendicular to the loop plane, then a shielding current, $I_{cir,b}=\mu_0 H_{\bot} A_b /L_b$, circulates in the loop to restore the initial flux\cite{mercereau}, where $A_b$ is the effective flux capture area of the loop in the bottom electrode and $L_b$ is the inductance of a thin-film elliptic (narrow) loop. What really matters is the magnetic flux $\Phi_b=\mu_0 H_{\bot} A_b$ applied to (although not threading) the loop; to a very high degree of approximation the capture area in this case is the loop inner area $A_b \approx \pi a b$. Applying the Ampere's circuital law to the (tangential) shielding current, $I_{cir,b}$, we easily derive the normal magnetic field, $H_{\nu,b} = I_{cir,b}/2 W_{b} =K_b H_{\bot}$, where $K_b=\mu_0 A_b/2W_{b} L_b$ is the conversion factor from the transverse to the barrier-parallel field\cite{SUST12}.

\noindent If we now consider the top/wiring electrode, for symmetry reasons and considering that the shielding current circulates on the opposite side of the barrier, we have that $H_{\nu,t} = -K_t H_{\bot}$. In the case of equal width annuli, then $K_b=K_t$ and the total normal field, $H_{\nu,b}+H_{\nu,t}$, is null meaning that the field generated by the currents circulating in the top electrode fully compensates that produced by the shielding currents induced in the bottom one. However, in general, $\Delta K=K_b-K_t \neq 0$ and the total normal field is uniform and proportional to the transverse field, $H_{\nu}=\Delta K\, H_{\bot}$; therefore, for short EAJTJs it gives rise to Fraunhofer-like threshold curves. To demonstrate this we resort to Eq.(\ref{gra}) in \elli coordinates, $(\partial \phi/\partial \nu, \partial \phi/ \partial \tau) $= $ \kappa a \mathcal{I}(\tau) (H_\tau, -H_\nu)$, which, in our one-dimensional approximation, yields $d\phi/d\tau= -\kappa a H_\nu \mathcal{I}(\tau)=-\kappa a \Delta K\, H_\bot \mathcal{I}(\tau)$. By integrating $d\phi/d\tau$ and introducing the dimensionless field $h_\bot\equiv\kappa a H_\bot$, we have:
\vskip -20pt
\begin{equation}
\label{phiditauisland}
\phi(\tau) = -h_{\bot} \Delta K\, \text{E}(\tau,e^2) + \phi_0.
\end{equation} 

\noindent Reiterating the calculation of Sec.IIC, with $\phi$ as in Eq.(\ref{phiditauisland}), we get the MDP of a short island-type EAJTJ in a transverse field:
\vskip -20pt
$$i_c(\pm h_{\bot})= \frac{1}{\text{E}(e^2)}\left| \int_{0}^{\pi/2} \!\!\!\!\! \mathcal{I}(\tau) \cos \left[ h_{\bot} \Delta K\, \text{E}(\tau,e^2)\right] d\tau \right|=\frac{1}{\text{E}(e^2)}\left| \int_{0}^{\text{E}(e^2)} \!\!\!\!\! \cos h_{\bot}z dz \right|=$$
\vskip -20pt
\begin{equation} \label{Icperpisland}
= \left| \frac{\sin h_{\bot} \Delta K\, {\text{E}(e^2)}}{h_{\bot} \Delta K\, {\text{E}(e^2)}} \right|=\left| \text{Sinc}\, h_{\bot} \Delta K\, {\text{E}(e^2)} \right|.
\end{equation}

\noindent Being its derivative is null, according to Eq.(\ref{psge}), a uniform normal field produces no effects on electrically long EAJTJs.

\subsection{Modified Lyngby-type geometry}

We now analyse the modified Lyngby-type geometry in which the left (right) side of the bottom (top) electrode is a semi-infinite plane; the bottom electrode of the configuration is shown in Figure~\ref{bottomelectrodes}(b). On the right side of the bottom electrode, the top/wiring electrode acts as a ground plane and squeezes the field lines generated by the screening currents in such a way that most of the magnetic energy is confined between the electrodes. The inductance per unit length of such superconducting microstrip transmission line\cite{chang79,PRB09} is $\mathcal{L}_b=\mu_0 \Lambda_b /W_{b}$, where $\Lambda_b$ is the bottom current penetration\cite{SUST13a,footnote} $\frac{\lambda_{b}}{2} \coth \frac{d_{b}}{2\lambda_{b}}+\frac{\lambda_{t}}{2} \tanh \frac{d_{t}}{2\lambda_{t}}$. We claim that the contribution to the loop inductance of the left side of the bottom electrode is negligibly small, therefore the loop inductance is given by the product of the ellipse mean semi-perimeter by the bottom inductance per unit length, $L_b=P \mathcal{L}_b/2$. This inductance is much smaller than that of a free standing elliptic loop used for the island-geometry\cite{vanDuzer}.

\noindent Being $W_{b}\!>>\!d_m$, for $0\leq\tau\leq\pi$, the normal magnetic field penetrating the barrier is\cite{meyers,SUST12} $H_{\nu,b} = I_{cir,b}/W_{b}= K_b H_{\bot}$, where now the field conversion factor\cite{PRB09} is $K_b=2 A_b/P \Lambda_b$. We note that $K_b$ is independent of $W_{b}$, since the circulating current linearly increases with the strip width. The shielding currents mostly flow along the outer electrode boundary, therefore, for $-\pi\leq\tau\leq0$, no normal field threads the \Jos barrier. If we now focus on the specularly symmetric top/wiring electrode, we have that $H_{\nu,t}= -K_t H_{\bot}$ for $-\pi\leq\tau\leq0$ and zero elsewhere. Here $K_t=2 A_t/P \Lambda_t$, with $\Lambda_t=\frac{\lambda_{t}}{2} \coth \frac{d_{t}}{2\lambda_{t}}+\frac{\lambda_{b}}{2} \tanh \frac{d_{b}}{2\lambda_{b}}=d_j-\Lambda_t$. Summing up the contributions from both electrodes, the total normal field is given by a $2\pi$-periodic step function having different amplitudes in the two half-periods:
\vskip -5pt
\begin{equation}
\label{cases}
\frac{H_{\nu}(\tau)}{H_{\bot}}=\begin{cases}{-K_t  }&\mbox{if } -\pi\leq\tau\leq0 \\
{K_b } & \mbox{if } 0\leq\tau\leq\pi. \end{cases} 
\end{equation}

\noindent Starting from Eq.(\ref{cases}),	it is not difficult to derive that for short \elli \juns with the modified Lyngby geometry in a transverse magnetic field:
\vskip -5pt
$$i_c(\pm h_{\bot})= \frac{1}{2}\left| \text{Sinc}\, \frac{\pi d_m \Phi_b}{ \Lambda_b \Phi_0 } +\text{Sinc}\, \frac{\pi d_m \Phi_t}{ \Lambda_t \Phi_0 } \right|,$$

\noindent which, in the case of equal fluxes, again reduces to a pure Fraunhofer pattern that only depends on the coupled flux (and not at all on the \jun ellipticity). It is worth to point out that, due to flux focusing effects, the effective capture area of this geometry is much larger than the ellipse inner area. As far as long \juns concerns, the normal field discontinuities at the ellipse poles result in two Dirac terms in the PSGE in Eq.(\ref{psge}), corresponding\cite{likharev,PRB10} to local extra conditions on the spatial phase derivative.

\subsection{Lyngby-type geometry}

\begin{figure}[tb]
\centering
\includegraphics[width=6cm]{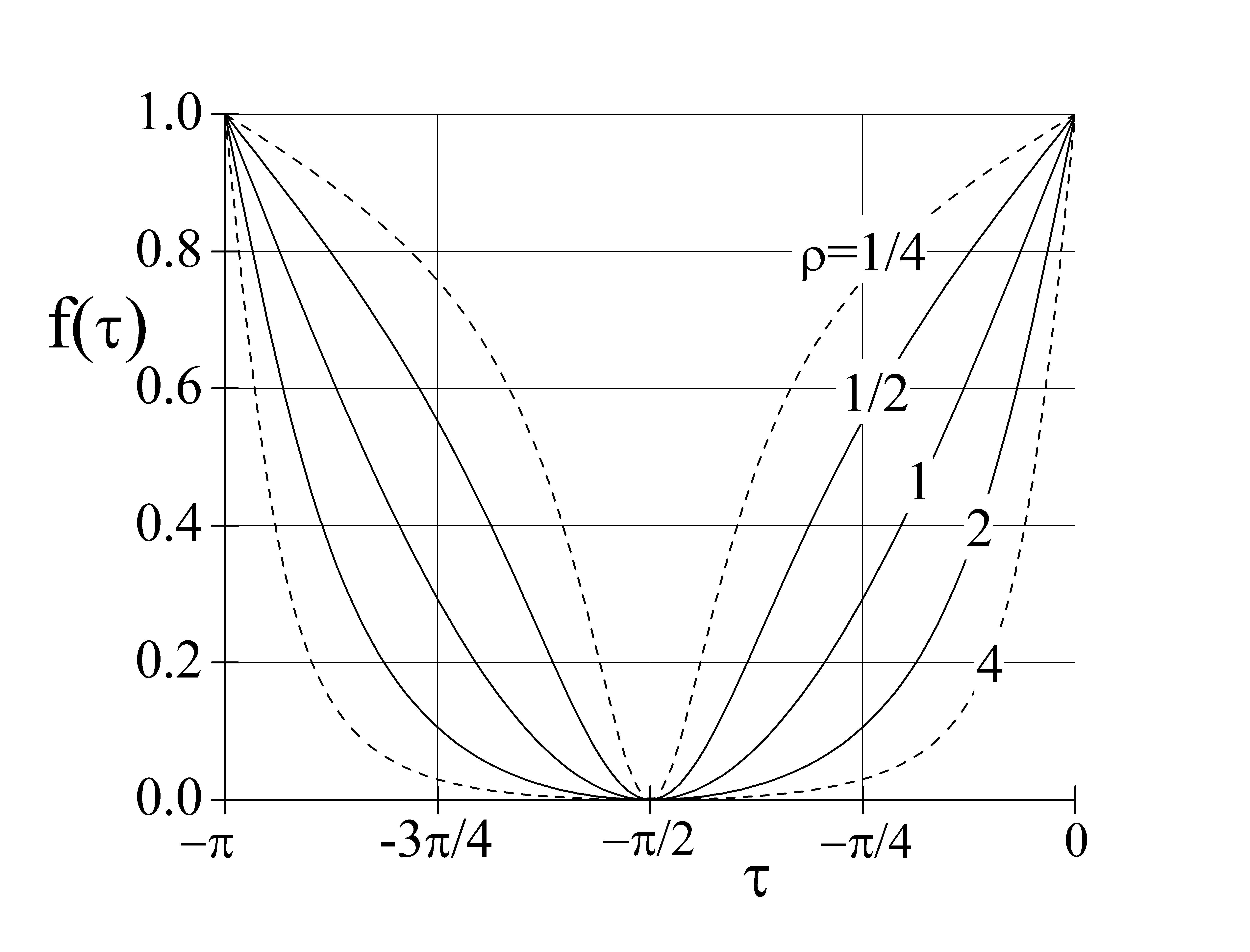}
\caption{Plot of $f(\rho,\tau)= 1+\rho\, \mathcal{I}^{-1}(\rho,\tau) \sin\tau$ for $-\pi\leq\tau \leq0$ and different $\rho$ values.}
\label{fvstau}
\end{figure}

Let us consider the bottom electrode of a Lyngby-type EAJTJ sketched in Figure~\ref{bottomelectrodes}(c). The loop width is constant and equal to $W_{b}$, on the right side and increases when we move to the left side. With respect to the previous geometry, this makes the transition to zero of the bottom normal field smoother as we cross the ellipse poles. The calculation of how this field vanishes when the circulating current moves away from the barrier is not an easy task, but, clearly this effect occurs faster for prolate junctions. Being $W_{b}\!<<\!b$, we will suppose that the field decay only depends on $\rho$ and can be qualitatively described by the function $f(\rho,\tau)= 
1+\rho\, \mathcal{I}^{-1}(\tau) \sin\tau$ plotted in Figure~\ref{fvstau} for different values of $\rho$ (in the range $-\pi\leq\tau\leq0$). Summarizing, the normal field, $H_{\nu,b}(\tau)$, generated by the screening currents circulating in the bottom electrode is given by the continuous expression:
\vskip -10pt
\begin{equation}
\label{bottom}
\frac{H_{\nu,b}(\tau)}{H_{\bot}}=\begin{cases}{K_b f(\rho,\tau)}&\mbox{if } -\pi\leq\tau\leq0 \\
{K_b } & \mbox{if } 0\leq\tau\leq\pi. \end{cases} 
\end{equation}
\noindent Analogously, for the specularly symmetric top electrode it is:
\vskip -10pt
\begin{equation}
\label{top}
\frac{H_{\nu,t}(\tau)}{H_{\bot}}=\begin{cases}{-K_t  }&\mbox{if } -\pi\leq\tau\leq0 \\
{-K_t f(\rho,-\tau)} & \mbox{if } 0\leq\tau\leq\pi. \end{cases} 
\end{equation}

\noindent Whatever are the electrode inductance, the conversion factors are\cite{SUST13b} $K_{b,t}=\mu_0 A_{b,t}/2W_{b,t} L_{b,t}$. The total normal field is $H_{\nu}=H_{\nu,b}+H_{\nu,t}$, and is given by the continuous function: 
\vskip -10pt
\begin{equation}
\label{total}
\frac{H_{\nu}(\tau)}{H_{\bot}}=\begin{cases}{ K_b f(\rho,\tau)-K_t }&\mbox{if } -\pi\leq\tau\leq0 \\
{K_b-K_t f(\rho,-\tau)}  & \mbox{if } 0\leq\tau\leq\pi. \end{cases} 
\end{equation}

\begin{figure}[tb]
\centering
\includegraphics[width=4cm]{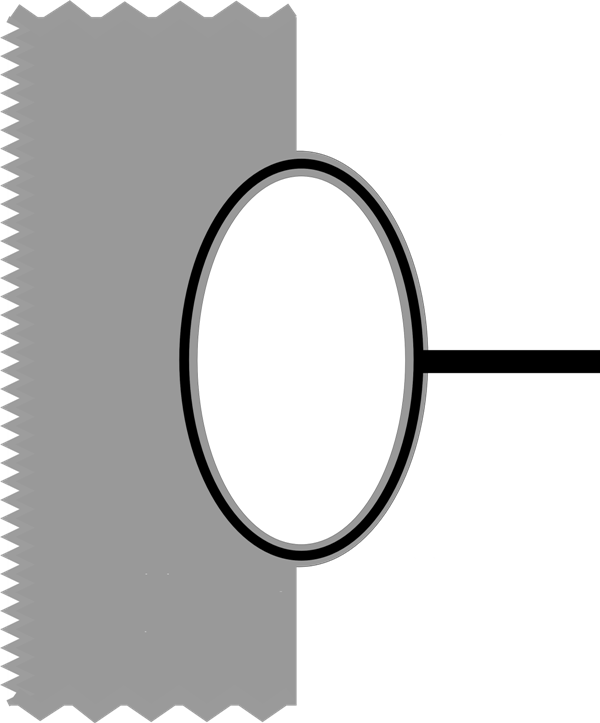}
\caption{Sketch of an \elli\ann\Jos\jun with \text{ratchet} geometry. The bottom electrode is in gray while the top/wiring electrode is black.}
\label{ratchet}
\end{figure}

\noindent In the case of symmetric electrodes ($K_{b}\simeq K_{t}=K$), the above expression greatly simplifies to $H_{\nu}(\tau)=H_{\bot} K \rho\, \mathcal{I}^{-1}(\tau) \sin\tau$ which is the normal component of a uniform in-plane field $K H_{\bot}$ applied in the direction of the $X$-axes, which is the direction of the current carrying electrodes. Remarkably, the critical current of both short and long Lyngby-type (symmetric) EAJTJs in a transverse field is expected to modulate exactly as described in Sections II and III for an in-plane magnetic field. This result is supported by magnetostatic simulations\cite{JAP08} showing that for Lyngby-type ring-shaped \juns in presence of a transverse field the normal magnetic field has a sinusoidal dependence on the polar angle $\theta$, independent of the annulus radius. Later on, it was experimentally proved\cite{JAP07} (although at that time not explained) that the field conversion factor $K$ of Lyngby-type ring-shaped \juns increases linearly with the mean radius, $r$; this is consistent with the capture area proportional to $r^2$ and an inductance proportional to $r$ (as in the case of the modified Lyngby geometry). Further support of the validity of Eq.(\ref{total}) will be provided by the experimental data reported in Section V. 

\noindent With $K_{b}\neq K_{t}$, Eq.(\ref{total}) results in an asymmetric field profile, $H_{\nu}(-\tau) \neq -H_{\nu}(\tau)$. The consequences resulting from the asymmetric boundary conditions imposed by a non-uniform external magnetic field at the extremities of both short and long linear Josephson junctions have recently been investigated; the field asymmetry is responsible for a degeneracy of the (extrapolated) critical field, $H_\parallel^c$, that was numerically demonstrated\cite{JAP10} and experimentally verified\cite{APL11}. Asymmetric field profiles in long \Jos junctions were recently looked for\cite{carapella,goldobin01} in order to exploit the rectifying property\cite{magnasco} of a ratchet potential. It is easy to recognize that the configuration depicted in Figure~\ref{ratchet} (bottom electrode in gray and top/wiring electrode in black) implements the ideal step-like deterministic ratchet potential with the normal field being constant on the left side ($K_t H_{\bot}$) and null on the right side. 

\vskip -15pt
\section{The measurements} 

\subsection{The samples and the experimental setup}

\begin{figure}[tb]
\centering
\subfigure[ ]{\includegraphics[height=2.5cm]{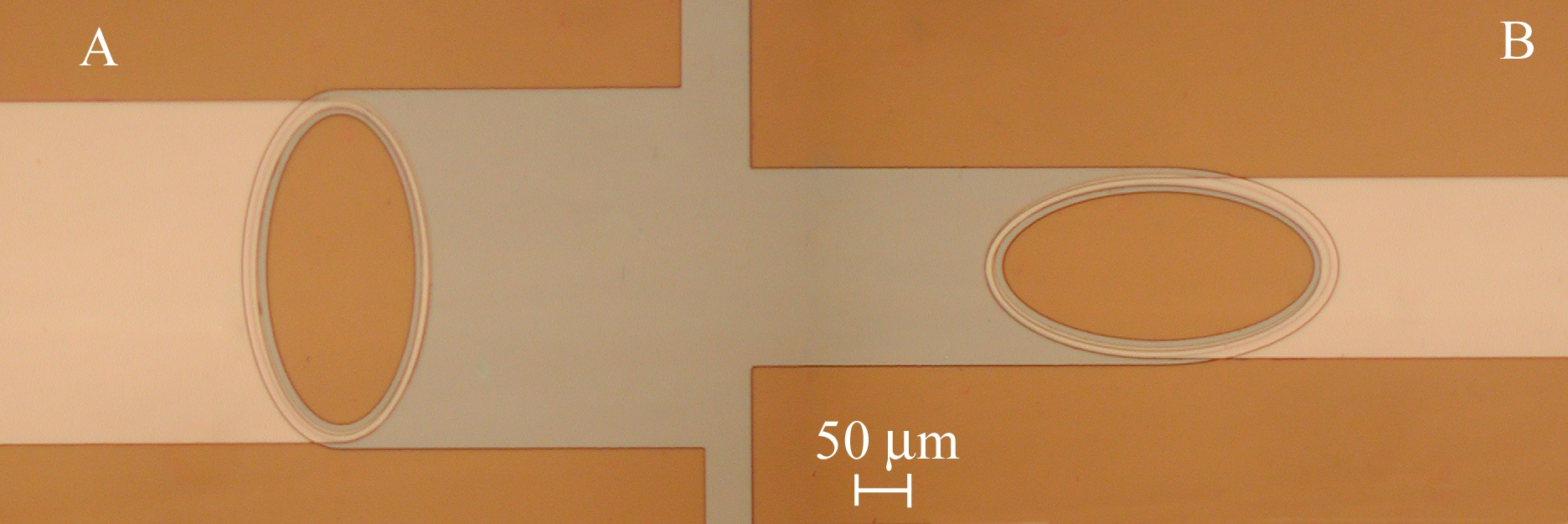}}
\subfigure[ ]{\includegraphics[height=4cm]{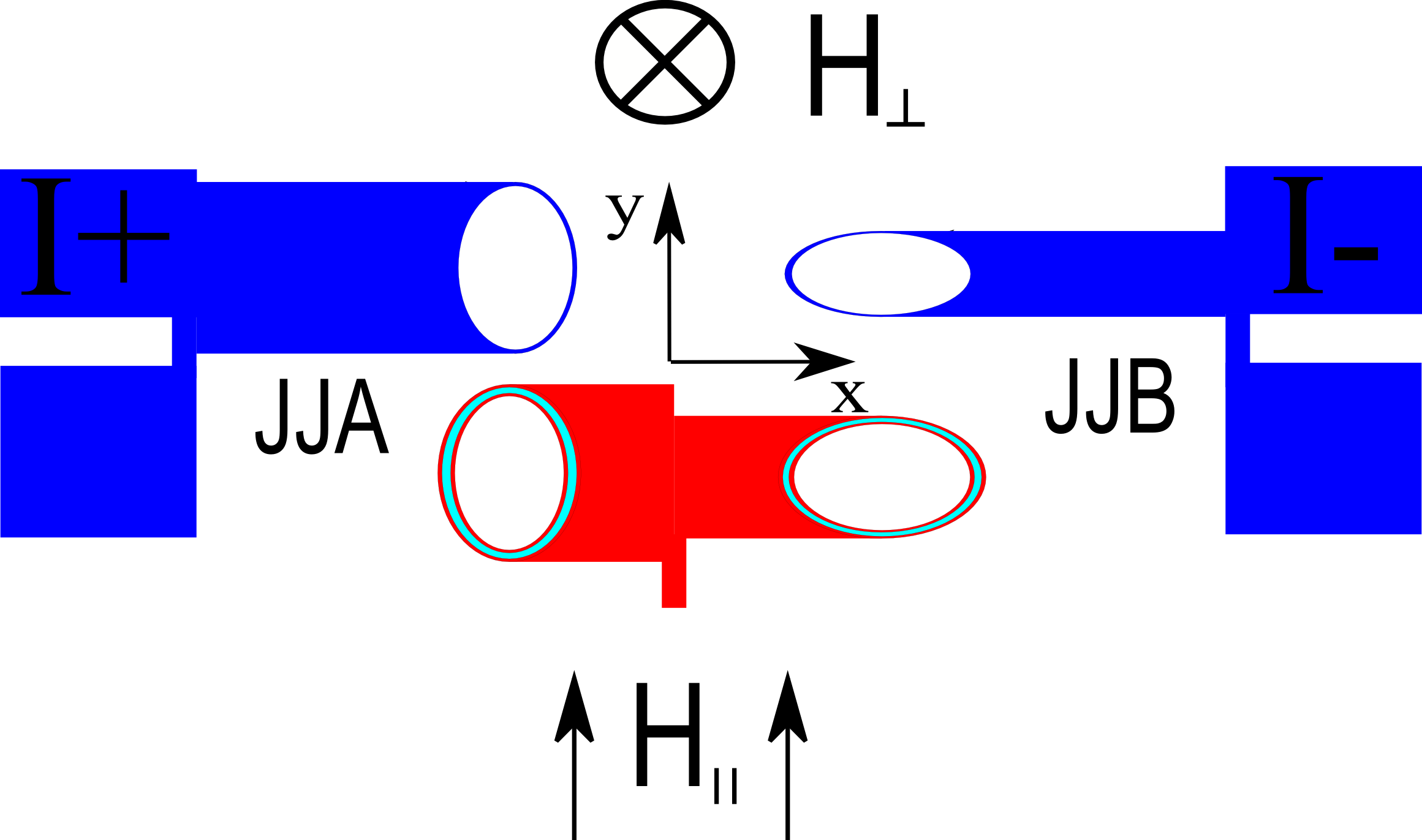}}
\caption{(Color online)(a) Optical image of one of our samples; (b) Exploded sketch of our layout; for clarity, the top/wiring layer (in blue) is shifted slightly upward with respect to the bottom layer (in red).}
\label{layout}
\end{figure}

Using the well known and reliable selective niobium anodization process\cite{snap} we have realized high-quality window-type $Nb/Al$-$AlOx/Nb$ EAJTJs. The details of the trilayer deposition and of the fabrication process can be found elsewhere\cite{granata07}. As shown by the photograph in Figure~\ref{layout}(a), two Lyngby-type \elli \ann junctions, named $A$ and $B$, were made having the same eccentricity, but rotated $90^o$ relative to one another. The \jun minor and major inner axes were, respectively, $140$ and $280\,\mu m$ and the nominal annulus width was $W=10\mu m$: however, due to different anodization conditions from wafer to wafer, the effective width was in the range of $7$-$9\,\mu m$. Figure~\ref{layout}(b) shows in more details the chip layout in which the elliptic tunnel barrier is sandwiched between two simply connected superconducting layers; the top/wiring layer (in blue) is shifted slightly upward with respect to the bottom layer (in red). 

\noindent Niobium anodic oxide ($80\, nm$ thick) and an extra dielectric layer made of rf-sputtered silicon dioxide ($120\, nm$ thick) provided the electrical insulation between the base electrode and the wiring film around the \jun area. This $200\, nm$ thick insulating by-layer outside the \elli ring, also called idle region, alters both the static and dynamic properties of the junctions; on linear one-dimensional long \juns it has been proved by both numerical simulation\cite{caputo} and experiments\cite{JAP95} that, as far as the static properties are concerned, the only effect of the idle region is to increase the magnetic energy stored in the fluxons, i.e., it introduces a scaling factor on the field strength. The different widths of the narrow part of the base and top electrodes, respectively, $W_{b}=20$ and $W_{t}=W=6\, \mu m$, result in a lack of full specular symmetry of our samples.  The different thicknesses of the base and top/wiring layer also contribute to the system asymmetry. For our samples the thicknesses of the base and top/wiring $Nb$ electrodes were, respectively, $d_b=200\,nm$ and $d_{t}=(40+500)\,nm \approx 6 \lambda_{Nb}$, with $\lambda_{Nb}(T=4.2K)=90\,nm$, resulting in magnetic and current penetration depths of, respectively, $d_m\approx \,162 nm$ and $d_j\approx \,182 nm \simeq 1.12 d_m$. Furthermore, the bottom and top current penetration depths were, respectively, $\Lambda_b\approx \,100 nm$ and $\Lambda_t = d_j-\Lambda_b \approx \,82 nm \simeq 0.82 \Lambda_b$. The critical current density of our samples was measured on electrically small cross-type \juns realized during the same deposition batch on different chips; at $T=4.2\, K$, we found $J_c=53\,A/cm^2$ corresponding to $\lambda_J\approx 52 \mu m$. Figure~\ref{selffield}(b) also shows that in our samples there is a $5\,\mu m$ wide idle region only on the outer left side of the junction; taking into account this asymmetric idle region, it is $\lambda_J\approx 58 \mu m$ on the \jun left side. For our calculation we will use the average value $\lambda_J\approx 55 \mu m$.

\noindent Our setup consisted of a cryoprobe inserted vertically in a commercial $LHe$ dewar. The cryoprobe was magnetically shielded by means of two concentric $Pb$ cans and a cryoperm one; in addition, the measurements were carried out in an rf-shielded room. The external magnetic field could be applied both in the chip plane or in the orthogonal direction. The chip was positioned in the center of a long superconducting cylindrical solenoid whose axis was along the $Y$-direction (see Figure~\ref{quarterellipse}) to provide an in-plane magnetic field, $H_{||}$. The transverse magnetic field, $H_\bot$, was applied by means of a superconducting cylindrical coil with its axis oriented along the $Z$-direction. All measurements were carried out at  $T=4.2$K.


\subsection{In-plane magnetic diffraction patterns} 

In this section we report the measurements carried out on \juns A and B of Figure~\ref{layout} having, respectively, $\rho_A=2$ and $\rho_B=0.5$, according to our notations (see Figure~\ref{quarterellipse}). For both \juns the mean perimeter is $P=4  \text{E}(1\!-\!\rho_B^2)\, a_B \approx 4\times 1.211 \times 145 \mu m \approx 700 \mu m$, i.e., much longer than the \Jos penetration depth, $\ell=P/\lambda_J\approx 13$ (it is $\text{E}(1\!-\!\rho_B^2)=\rho_B \text{E}(1\!-\!\rho_A^2)\approx 1.21$). A large number of such samples were investigated and they all showed not only the zero-field critical current, $I_{c,0}$, but also the maximum critical current, $I_{c,max}$, considerably smaller than about the $70\%$ of the current jump at the gap voltage, $\Delta I_g$, typical of short $Nb/Al$-$AlOx$-$Al/Nb$ junctions. As anticipated in Section IIIC, this is the first signature of a non-uniform bias current distribution and, more importantly, of the self-field effects.

\begin{figure}[tb]
\centering
\subfigure[ ]{\includegraphics[width=7cm]{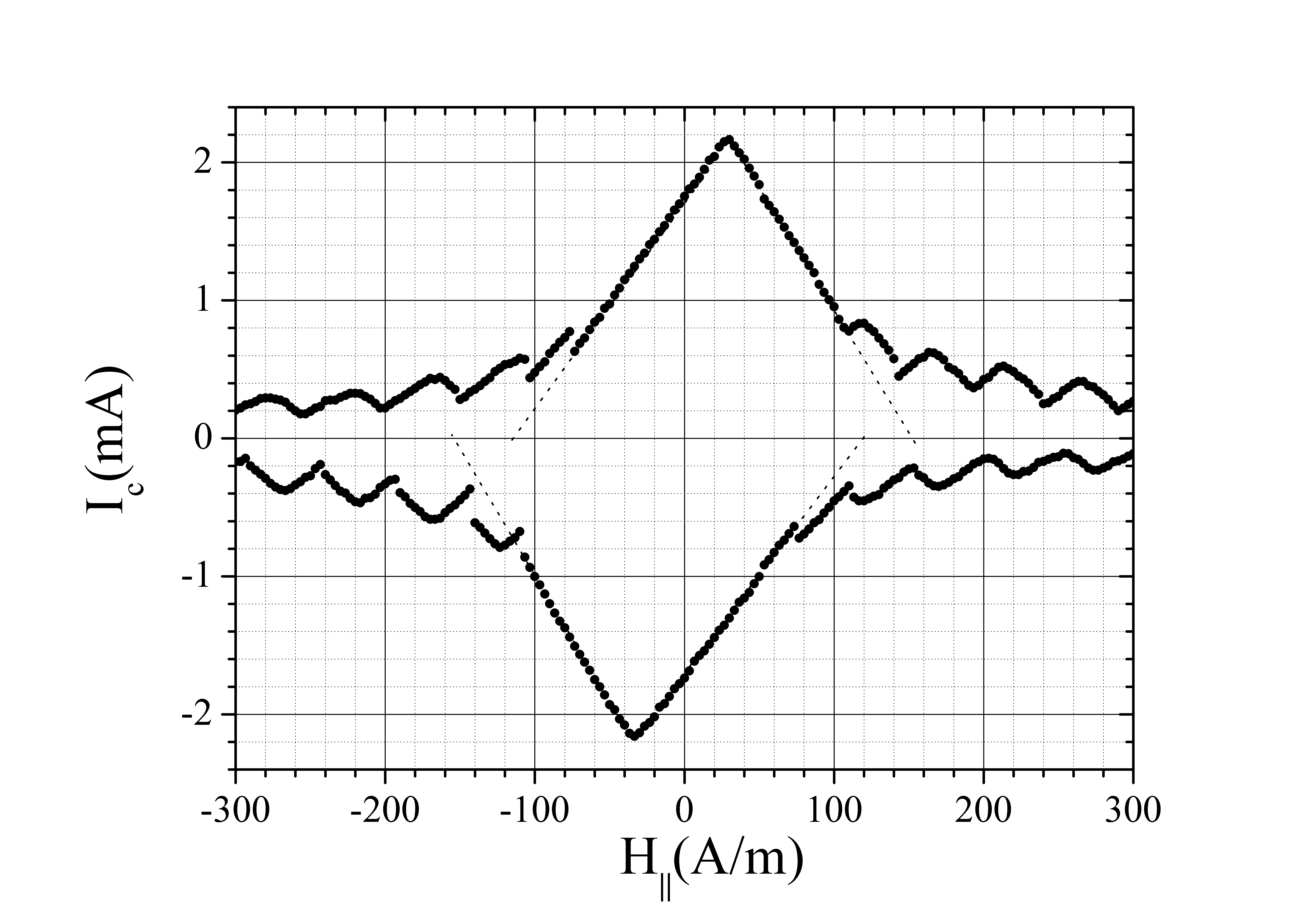}}
\subfigure[ ]{\includegraphics[width=7cm]{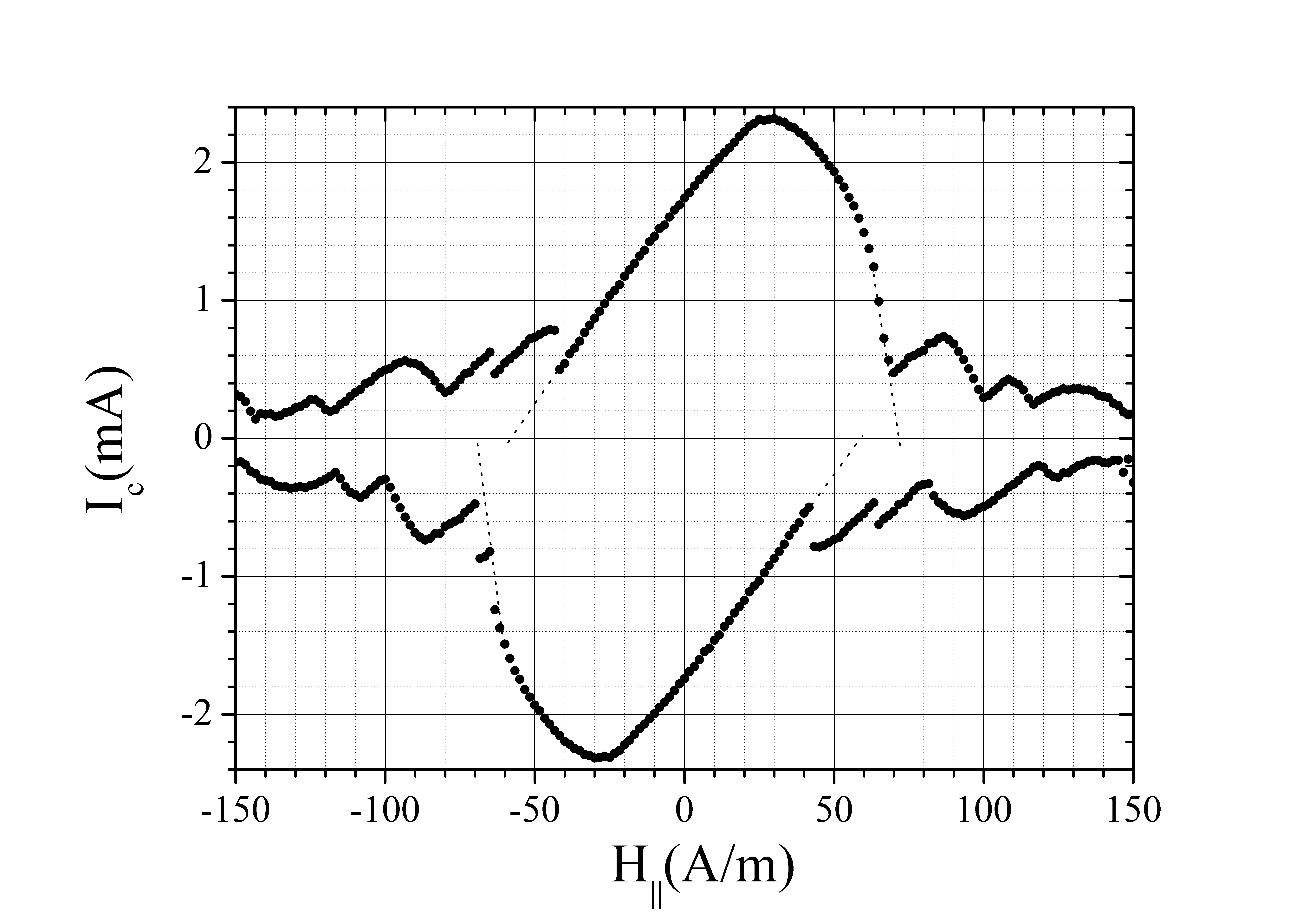}}
\caption{Experimental magnetic diffraction patterns of our EAJTJs with an in-plane magnetic field: (a) \jun A and (b) \jun B. The extrapolated dotted lines help to locate the critical fields.}
\label{par}
\end{figure}

\noindent Figures~\ref{par}(a)-\ref{par}(b) display the MDPs, respectively, of samples A ($\Delta I_g=5.7\,mA$) and B ($\Delta I_g=5.5\,mA$) with an in-plane magnetic field applied in the direction perpendicular to the bias current flow. At a first glance, we observe that both datasets are quite different from what is expected for a long EAJTJ with a similar normalized length (see Figure~\ref{2and4Pi}(b)). With real devices, the measurements of maximum supercurrent as a function of the external field often yield the envelop of the lobes, i.e., the current distribution switches automatically to the mode which for a given field carries the largest supercurrent. Sometimes, for a given applied field, multiple solutions are observed by sweeping the \jun current-voltage characteristic many times. Nevertheless, the (first) critical field $H^c_\parallel$ can still be obtained extrapolating to zero the MDP first lobe (see dotted line in Figures~\ref{par}(a)-\ref{par}(b)). Furthermore, while all the theoretical threshold curves derived in the previous sections were implicitly meant to be symmetric with respect to the bias current \textit{and/or} magnetic field inversion, the experimental MDPs only retain the symmetry with respect to the simultaneous inversion of the bias current \textit{and} of the magnetic field. The pattern skewness is mainly to be ascribed to the $Y$-component of the normal self-field; in the second and fourth MDP quadrants this component adds to the external field, while in the first and third quadrants it partially compensates the applied field; this results in different slopes of the first lobe and therefore in different (extrapolated) critical fields. Our data are consistent with the postulate in Section IIIC, of the current-perpendicular self-field being larger for prolate EAJTJs. In different words, the in-plane MDP of our samples would not have been skewed, if the magnetic field were been applied in the direction of the bias current. Furthermore, from the numerical simulations of Figure~\ref{2and4Pi}(b) we expected the critical fields of \juns A and B to be, respectively, $\approx 5J_c \lambda_J \simeq 145\,A/m$ and $\approx 2.4J_c \lambda_J \simeq 70\,A/m$ which are close to the measured average values (respectively, $130$ and $65\,A/m$). The larger $I_{c,max}/ \Delta I_g$ ratio observed for \jun B, can be explained by smaller current-parallel self-fields in oblate rather than prolate EAJTJs, but is also consistent with the supposition that, as the height of the current carrying electrodes gets smaller, the current density profile becomes more uniform.


\subsection{Transverse magnetic diffraction patterns}

\begin{figure}[tb]
\centering
\subfigure[ ]{\includegraphics[width=7cm]{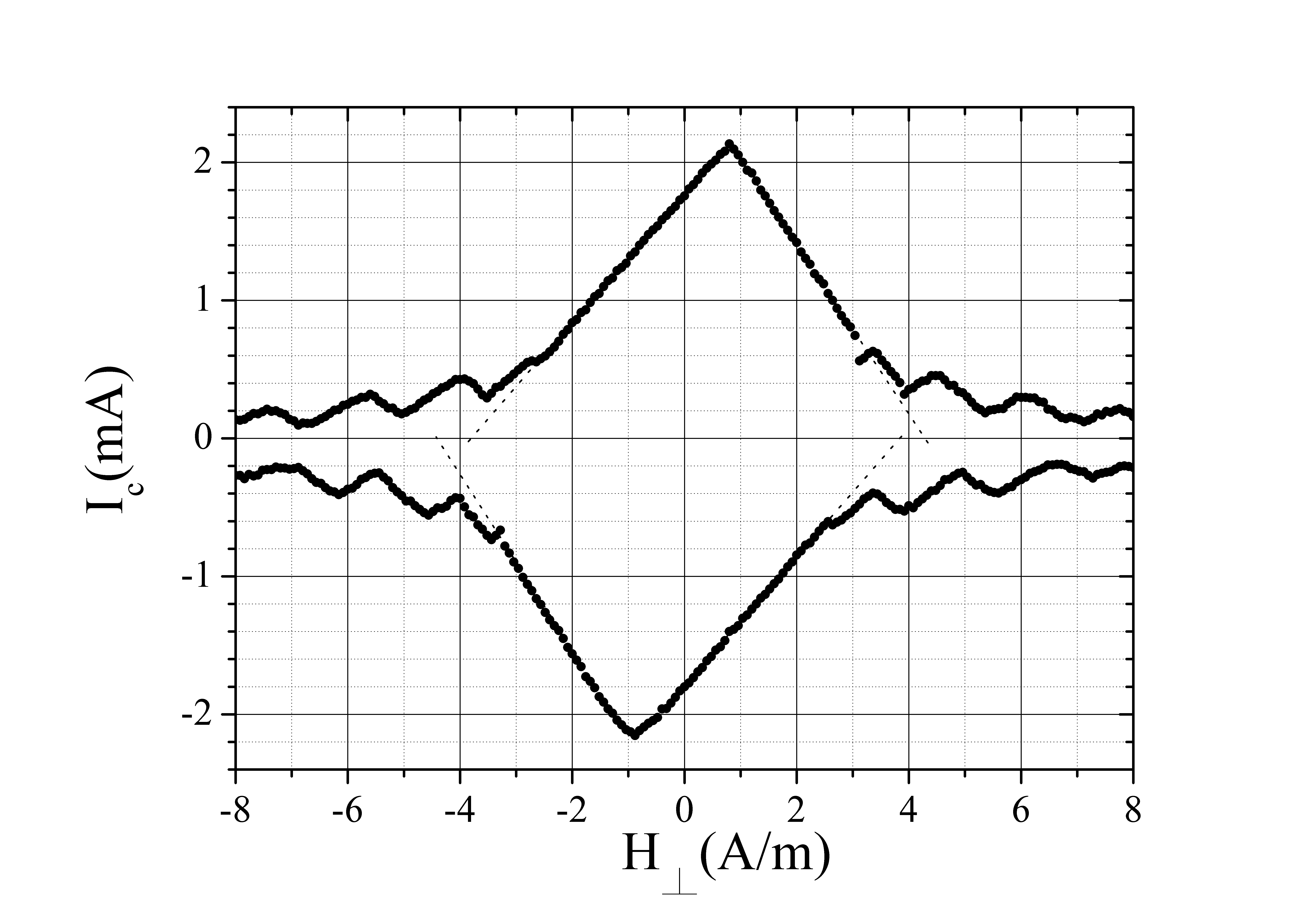}}
\subfigure[ ]{\includegraphics[width=7cm]{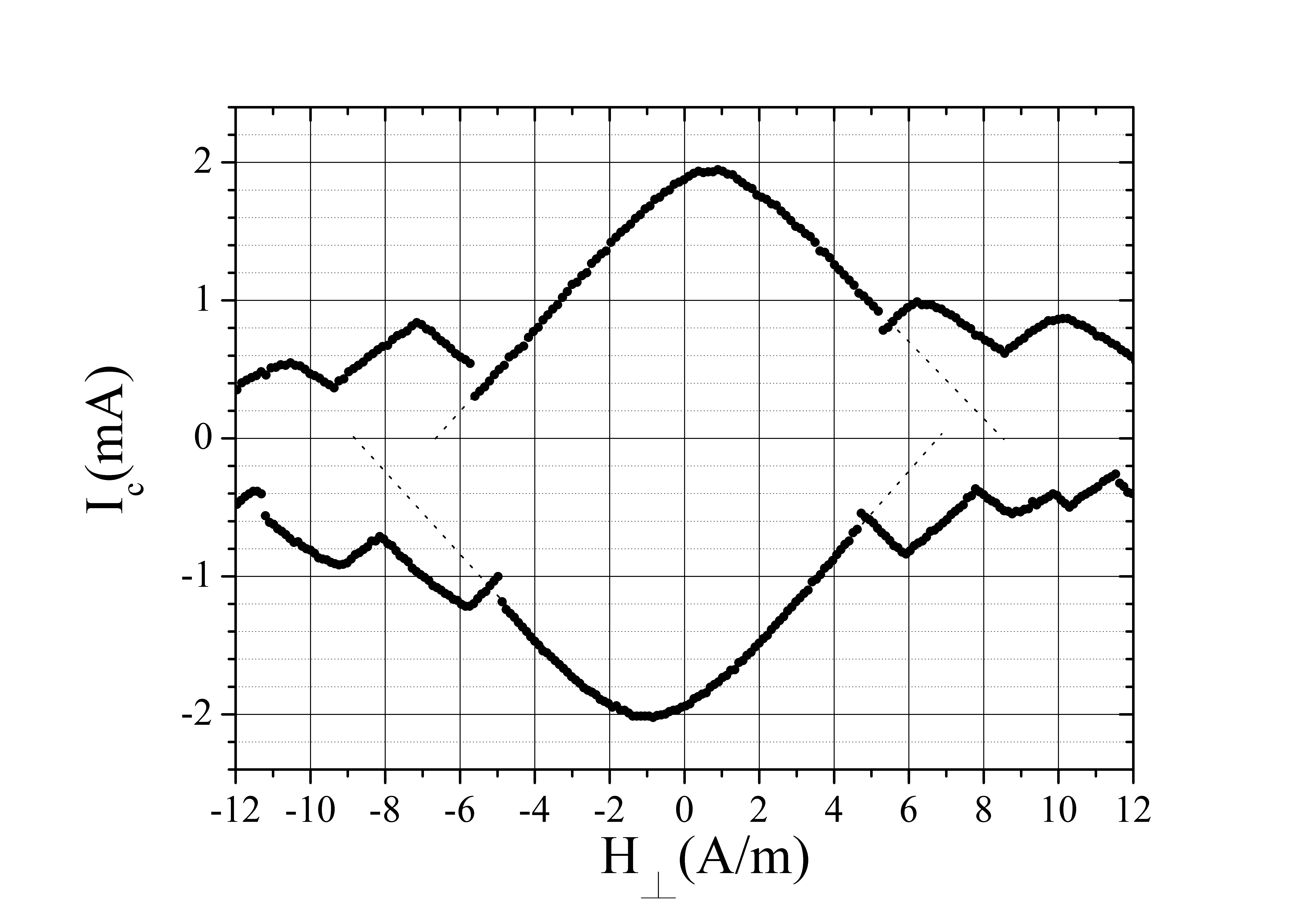}}
\caption{Experimental threshold curves of our EAJTJs in a transverse magnetic field: (a) \jun A and (b) \jun B. The dotted lines help to locate the critical fields.}
\label{perp}
\end{figure}

Figures~\ref{perp}(a)-\ref{perp}(b) display the $I_c$ vs. $H_\bot$ dependences of, respectively, the same \juns A and B reported in Figures~\ref{par}(a)-\ref{par}(b).
For sample A we observe that the transverse MDP is practically indistinguishable from its in-plane counterpart (apart from a field factor scale); this confirms our finding of Section IV that for Lyngby-type annuli a transverse field is equivalent to an-in plane field applied in the direction of the current flow and at same time substantiate the presence of a current-parallel self-field (otherwise the transverse pattern would not be skewed). The comparison of the transverse and in-plane MDPs of sample B further supports this picture, if we recall that the effect of the current-parallel self-field is reduced, if not negligible, in oblate junctions. Now the transverse critical field is larger for \jun B ($7.5A/m$), rather than \jun A ($4.1A/m$); this was expected considering that a $90^o$ rotation of the in-plane field corresponds to a transformation $\rho \to 1/\rho$. In other words, in presence of a transverse magnetic field, the system symmetry is broken along the direction of the current flow. This makes sample A about 32 times more sensitive to $H_\bot$ than to $H_\parallel$, the field conversion factor being about $8.6$ for \jun B. 

\subsection{Vortex trapping}

It was possible to trap Josephson static vortices (fluxons) on a statistical basis by means of fast coolings of the samples through their superconducting transition. The trapping probability is know to grow with the speed of the normal-to-superconducting transition\cite{PRB06,PRB08}.  After a successful trapping procedure the zero-voltage critical current is considerably smaller and a stable finite-voltage current branch, called zero-field step, appears in the junction current-voltage characteristic indicating that the bias current forces a single fluxon to travel along the ellipse perimeter in the absence of collisions. The corresponding flux quantum must be trapped in the superconducting loop formed by either the bottom or top electrode. For our samples the current branch associated with one fluxon had an amplitude larger than $1\,mA$ and an asymptotic voltage $V_1 \approx 29\,\mu V$ which results in an average speed, $P V_1/ \Phi_0\approx 10^7\, m/s$, considerably smaller than the Swihart velocity, $1.5 \times 10^7\, m/s$, typical of all-$Nb$ junctions evidencing, once again, indicating that the fluxon travels in the periodic potential\cite{PRB98} generated by the bias current. The dynamic properties of EAJTJs involves the nucleation, propagation and interaction of more than one fluxon or fluxon-antifluxon pair and will be discussed in a future work.

\section{Conclusions}

The static properties of one-dimensional \elli \ann \Jos tunnel \juns have been investigated theoretically and experimentally. Both short and long \juns were considered in the presence of an in-plane as well as a transverse magnetic field. For short \ann\juns we derived and computed the dependence of the critical current on a uniform in-plane magnetic field; it is found that the $I_c$ vs. $H$ dependence is determined by the sample ellipticity, although the first critical field can only change by 50\%. Different geometrical configurations have been considered for studying the effect of a magnetic field applied perpendicular to the ellipse plane; we established that for a Lyngby-type EAJTJ a transverse field emulates an in-plane field applied along the direction of the current flow. Further, we derived the proper perturbed sine-Gordon equation to describe both the statics and the dynamics of the phase difference across the barrier of an EAJTJ. The static solutions of the partial differential equation were numerically computed and we found that, for a given field, different phase profiles are possible depending on the number of static fluxon-antifluxon pairs nucleated at the ellipse poles where the derivative of the normal field is largest. We also evaluated the static self-field effect in long EAJTJs. Two planar structures characterized by not simply-connected electrodes have been considered in the experiments. Experimental data on high-quality long $Nb/Al$-$AlOx$-$Al/Nb$ EAJTs basically confirm the numerical predictions, provided the effects of the current limiting static self-field are taken into account. A transverse magnetic field is demonstrated to be several times more efficient than an in-plane one to modulate the junction critical current. For a given inner area, prolate EAJTJs (for which the inversion of the periodic potential occurs faster) are more efficient for the in-plane to transverse field conversion.

\noindent Unbiased elliptic annular junctions inherently have specular symmetry with respect to their principal axes:  quite obviously an in-plane magnetic field breaks the system symmetry along its direction. In this paper we have demonstrated that a transverse field breaks the symmetry along the direction of the current carrying leads; furthermore, in long EAJTJs the bias current itself also generates non-symmetric conditions, the asymmetry being more pronounced along the current direction for oblate ellipses and vice versa. Among other things, we have also suggested a simple geometrical configuration in which the magnetic field coupled to the \elli barrier lacks reflection symmetry, so accomplishing a nearly ideal rectifying potential in which a soliton is accelerated only in one half of the junction perimeter. The soliton dynamics in EAJTJs will be treated in a forthcoming paper.

\vskip -10pt
\section{Acknowledgments}
%
\noindent RM acknowledges the support of the Italian Consiglio Nazionale delle Ricerche under the Short Term Mobility Program 2014. RM and JM acknowledge the support from the Danish Council for Strategic Research under the program EXMAD.


\begin{thebibliography}{99}
\bibitem{note2} For \Jos \juns fabricated with the tri-layer technique, usually $d_t>4\lambda_t$, so $d_j/d_m \simeq \coth d_b / \lambda_{b} \coth d_b /2 \lambda_{b}$; with $d_b =2 \lambda_{b}$, it is $d_j/d_m \simeq 1.36$.
\end{thebibliography}
\end{document}